\newif\ifconfver
\confverfalse      

\ifconfver
    \documentclass[10pt,journal]{IEEEtran}
\else
    \documentclass[11pt,draftclsnofoot,onecolumn]{IEEEtran}
\fi
\date{}
\usepackage{xspace,empheq,fancybox,amsmath,amssymb,epsfig,subfigure,syntonly,times,amsthm}
\usepackage{psfrag,color,bm}
\usepackage{url,cite,footnote,xspace,syntonly,algorithm,algorithmic}
\usepackage{verbatim}
\usepackage{amsmath,amsthm}
\usepackage{amssymb}
\usepackage{amsfonts}
\usepackage{epsfig}
\usepackage{cite,footnote,xspace,syntonly,algorithm,algorithmic}
\usepackage{bm,empheq,fancybox,url}
\usepackage[letterpaper,top=1in,bottom=1in,left=1in,right=1in,bindingoffset=0in]{geometry}

\allowdisplaybreaks[4]

\input{mysymbol.sty}

\newcommand{\tr}{\mathrm{Tr}}
\newcommand*\widefbox[1]{\ovalbox{\hspace{1em}#1\hspace{1em}}}
\newcommand{\ind}{1\hspace{-1.5mm}1}
\newcommand{\ude}{{\underline{e}}}
\newcommand{\udv}{{\underline{v}}}
\newcommand{\udE}{{\underline{\ccalE}}}
\newcommand{\udF}{{\underline{\ccalF}}}
\newcommand{\udU}{{\underline{\ccalU}}}
\newcommand{\udV}{{\underline{\ccalV}}}
\newcommand{\udS}{{\underline{\ccalS}}}

\newcommand{\udA}{{\underline{\ccalA}}}

\begin{document}
\date{}

\title{Precoder Design   for Physical Layer Multicasting}
\author{
                \IEEEauthorblockN{{\it Hao Zhu \authorrefmark{1}}},
                \IEEEauthorblockN{{\it Narayan Prasad \authorrefmark{2}}},
                 and
                \IEEEauthorblockN{{\it Sampath Rangarajan \authorrefmark{2}}}

\vspace{0.5cm}
\thanks{\protect\rule{0pt}{1.5em}
\authorrefmark{1} Department of Electrical and Computer Engineering, University of Minnesota  (email: zhuh@umn.edu).}

\thanks{\protect\rule{0pt}{1.5em}
\authorrefmark{2} Mobile Communications and Networking Research, NEC Labs, Princeton (emails: \{prasad, sampath\}@nec-labs.com).}

\vspace{-1cm}
{\small
\begin{center} \[ \begin{array}{rl}
\text{\bf Submitted:} \quad \text{\today} \\
      \end{array}  \] \end{center}  }
 \vspace{-1cm}
}
\markboth{IEEE TRANSACTIONS ON SIGNAL PROCESSING (SUBMITTED)}%
{Zhu \MakeLowercase{\textit{et al.}}: Precoder Design for Physical Layer Multicasting}
\date{}
\maketitle
\thispagestyle{empty}
\pagenumbering{arabic}
\date{}
%
\begin{abstract}

\noindent This paper studies the  \emph{instantaneous rate} maximization and the \emph{weighted sum delay} minimization problems over a $K$-user multicast channel, where multiple antennas are available at the  transmitter as well as at all the receivers. Motivated by  the degree of freedom optimality and the simplicity offered by linear precoding schemes, we consider the design of linear precoders using the aforementioned two criteria. We first consider the scenario wherein the linear precoder can be any complex-valued matrix subject to rank and power constraints. We propose   cyclic alternating ascent based precoder design algorithms and establish their convergence to respective stationary points. Simulation results reveal that our proposed algorithms considerably outperform known competing solutions.
We then consider a scenario in which the linear precoder can be formed by selecting and concatenating precoders from a given finite codebook of  precoding matrices, subject to  rank and power constraints. We show that under this scenario, the instantaneous rate  maximization problem  is equivalent to a robust submodular maximization problem which is strongly NP hard. We propose a deterministic approximation algorithm and show that it yields a bicriteria approximation. For the weighted sum delay minimization problem  we propose a simple deterministic greedy algorithm, which at each step entails approximately maximizing a submodular set function subject to multiple knapsack constraints, and establish its performance guarantee. 
\end{abstract}

\smallskip


\normalsize{}
\newpage


\section{Introduction}\label{sec:intro}

Next generation wireless networks will require a spectrally efficient physical layer multicasting scheme in order to cater to important emerging applications such as real-time video broadcast, wherein a common information needs to be simultaneously transmitted to multiple users.  The design of spectrally efficient physical layer multicasting schemes via \emph{instantaneous rate maximization} has consequently received significant recent attention.
The seminal work of \cite{sidiro06} considers the design of the instantaneous rate maximizing  transmit beamforming (a.k.a. rank-1 linear precoding) scheme for multicast and proves it to be an NP-hard problem.
Efficient albeit sub-optimal designs of transmit beamforming (or equivalent rank-1 transmission schemes) for multicast have thus been proposed in \cite{sidiro06,jwangMC}. In addition, a hidden convexity of the multicast beamforming problem under certain channel conditions has been recently discovered in \cite{amir}. Another approach for designing beamforming vectors for multicast has been adopted in \cite{sstan}. In particular, \cite{sstan} assumes that users have been partitioned into non-overlapping user groups and then proceeds to design beam vectors (one for each group) and their power levels. Several efficient heuristics are suggested. This approach is further pursued in \cite{senaratne}, where formation of groups is also considered and transmissions pertaining to different groups are made orthogonal. Long-term beamforming
for scenarios where instantaneous channel state   is unavailable at the transmitter has been
addressed in \cite{lozano}. 
On the other hand, the optimal (i.e., instantaneous rate maximizing) linear precoding based multicasting scheme without rank constraints can be obtained via convex optimization \cite{jindal_isit06}. The scaling results derived in \cite{jindal_isit06} reveal that higher rank precoding is beneficial in the ubiquitous regime in which the number of users is larger than the number of transmit antennas. Indeed in this regime an open loop scheme with identity matrix precoder (whose size is equal to the number of transmit antennas) is asymptotically optimal.

This paper intends to address the main issue with such higher rank precoding for multicast, which is the increase in the decoding complexity at each user, particularly when the rank exceeds the number of its receive antennas. In particular, we consider the problem of designing  linear precoders for multicast subject to a given rank constraint, which allows us to address the trade off between spectral efficiency  and  decoding complexity. Compared to an existing recursive design based approach for constructing linear precoders for multicast \cite{kim_rws08} (see also \cite{kim_comm}) which can also accommodate an input rank constraint, our approach introduces auxiliary variables to reformulate the optimization problem and uses an alternating optimization method \cite{tseng01jota} to achieve a Karush-Kuhn-Tucker (KKT) stationary point. We note that an antenna subset selection scheme for multicast, which selects the optimal transmit antenna subset of a given size (assuming identity matrix precoder of that size), has been analyzed in \cite{parkD}.
Furthermore,  alternating optimization based algorithms have been proposed for several multicast precoder design algorithms in \cite{shi_ciss08} all of which involve the decoding mean squared error. Here we consider the achievable rate instead, which involves introducing more auxiliary variables in the alternating optimization and the resulting proof of convergence is also different.

In addition to the transmission rank constraint,  in certain practical systems  each user needs to be explicitly signaled about the choice of the  precoder employed by the transmitter, thus necessitating the choice to lie in a finite codebook. Instead of considering an optimal albeit unstructured finite codebook design, we focus on a more practical setup entailing a lower memory footprint and signaling overhead, wherein a higher rank precoder is constructed by concatenating codewords from a given (base) codebook of precoding matrices. Under this scenario we show that the instantaneous rate maximization problem falls in the realm of the robust submodular optimization \cite{krause_jmlr08} and is strongly NP-hard. We propose a deterministic approximation algorithm and show that it yields a bicriteria approximation.

Another  precoder design metric of interest for physical layer multicasting is the \emph{weighted sum delay}. The pertinent delay for each user is defined as the number of time intervals needed to accumulate enough information for  decoding a common message; and the weight assigned to a user is  determined by its priority in the multicasting system. Linear precoder design to minimize the weighted sum delay is considered under  rank and power constraints as well as under a finite codebook-constraint, for which the alternating optimization   and the submodularity, respectively, again become instrumental to develop efficient algorithms. We note that sum delay minimization over a discrete codebook has been recently considered in \cite{ZhangInfo}. However, the innovative algorithms designed in \cite{ZhangInfo} are based on an assumption (which holds for strongly LOS channels) that each user can receive its data from only one beamforming vector in the codebook and that all other vectors are essentially in the null space of that user's channel, i.e. transmission along any such vector will result in a negligible received signal strength at the user. In contrast, we make no such assumption and indeed allow each user to accumulate its useful signal across several intervals (where one or more precoders are employed for transmission in each interval) till it meets a threshold for reliable decoding.

The rest of the paper is organized as follows. Section  \ref{sec:PS}
presents the system model and formulates the two aforementioned precoder design problems. Efficient algorithms for maximizing the instantaneous rate  are developed in Section \ref{sec:maxinst}; while Section \ref{sec:mindelay} switches to the weighted sum delay minimization problem. The proposed algorithms are
tested and compared numerically to other known approaches in Section \ref{sec:sim} and the conclusions are presented in Section \ref{sec:conclusions}.

\vspace{1mm} \noindent {\it Notation:} Upper (lower) boldface
letters will be used for matrices (vectors); $(\cdot)^\dag$ denotes the complex-conjugate transposition; 
$\tr(\cdot)$ the matrix trace; rank$(\cdot)$ the matrix rank; $\mathbf 0$ the all-zero matrix;  
$\|\cdot\|_F$ the matrix Frobenious norm; and $|\cdot|$ the cardinality of a set as well as the determinant of a square matrix.


\section{System Model and Problem Statement}\label{sec:PS}


We consider a  MIMO wireless physical layer multicasting system consisting of a   base station (BS) equipped with $M$ transmit antennas and $K$ users, where the $k^{th}$ user is equipped with $N_k$ receive antennas  for $k = 1,\ldots, K$. All the $K$ users receive  common information from the BS. We let $\bbx^\tau \in \mathbb C^M$ denote the  signal vector transmitted by the BS on slot $\tau   \in \mathbb Z_+$,  where a slot denotes a resource unit in the code, frequency or time domain. Further, let $\bby_k^\tau \in \mathbb C^{N_k}$ be the   signal vector  received by user $k = 1,\ldots, K$ on slot $\tau$. Then, the input-output (I/O) relationship for the $k$-th user is modeled as
\begin{align}
\bby_k^\tau = \bbH_k^\tau \bbx^\tau + \bbz_k^\tau,~~\forall k \label{kuser_io}
\end{align}
where $\bbH_k^\tau\in \mathbb C^{N_k \times M}$ is the channel matrix that models the channel seen by the $k$-th user from the BS  on slot  $\tau$, and $\bbz_k^\tau \in \mathbb C^{N_k}$ is the additive complex Gaussian noise vector at the $k$-th user. The noise vectors are assumed to be mutually independent (across slots) complex Gaussian vectors and without loss of generality (Wlog)  they are each assumed to be white, i.e., $\bbz_k^\tau \sim \ccalC\ccalN(\bb0, \bbI)$. This is possible  via a whitening filter which can be absorbed into the channel matrix $\bbH_k^\tau$. A power budget is imposed on the transmitted signal as $\mathbb E [\|\bbx^\tau\|^2] \leq P$, $\forall \tau \in \mathbb Z_+$. It is further assumed that estimates of all the channel matrices $\{\bbH_k^\tau\}$ in \eqref{kuser_io} are available   at the BS, possibly by exploiting  reciprocity or feedback. In this paper for simplicity we assume that error free estimates are available to the BS.  Nonetheless,
the design methods presented in the sequel can be generalized to the scenario where only imperfect channel estimates are available. For example, one approach is to mimic
the naive zero-forcing based precoding design for multiuser MIMO and let the BS design the precoders after assuming the channel estimates available to it to be perfect. Another more sophisticated approach is also possible by explicitly modeling the CSI errors; see for instance \cite{vucic2009,jose_nectr10}.

Next, consider a simple communication scheme that uses linear transmit precoding at the BS.
To this end, suppose $d$ symbol streams are simultaneously transmitted by the BS on each slot and let $\bbs^{\tau} \in \mathbb C^d$ denote the coded and modulated symbol vector with $\bbW^{\tau} \in \mathbb C^{M\times d}$ denoting the corresponding precoding matrix.  Thus, the transmitted signal at the BS becomes $\bbx^\tau = \bbW^\tau \bbs^\tau$, and the Input/Output relationship per user $k$ is given by
\begin{align}
\bby_k^\tau = \bbH_k^\tau \bbW^\tau \bbs^\tau+ \bbz_k^\tau,~~\forall k. \label{kuser_ioc}
\end{align}
Wlog the encoded symbol vector is assumed to satisfy $E[\bbs^\tau \bbs^{\tau\;\dag}]=\bbI$. Therefore, the achievable rate at the $k$-th user for the scheme \eqref{kuser_ioc} can be expressed as
\begin{align}
R_k^\tau(\bbW^\tau) = \log \left|\bbI +  \bbH_k^\tau \bbW^\tau \bbW^{\tau\;\dag} \bbH_k^{\tau\;\dag}\right|, ~~\forall k. \label{Rkt}
\end{align}
Then, given the multicast system \eqref{kuser_ioc} and some prescribed precoder codebook  $\ccalC$  (as detailed later),  we are interested in the problem of selecting the precoding matrix $\bbW^\tau \in \ccalC$ under the following two goals. The first design criterion is to achieve the best instantaneous throughput on each slot $\tau$, or equivalently maximize the minimum of the rates $\{R_k^\tau\}$ among all the $K$ users. For simplicity, the slot index $\tau$ can be omitted under this scenario, and the problem of interest becomes
\begin{empheq}[box=\widefbox]{align} \label{p1}
(P1)~~~~~\max_{\bbW\in\ccalC} ~~ \min_{k=1,\ldots,K} ~~R_k (\bbW). ~~~~~
\end{empheq}
%
Clearly, the precoder design problem (P1) focuses on the instantaneous throughput at each channel use that  can be achievable for all users. In some circumstances, it is more meaningful to look at a weighted average performance across all the $K$ users. Here, we consider a quasi-static fading scenario where in each scheduling interval (defined over the time domain) the BS repeatedly transmits the same message over $L$ orthogonal slots. The BS continues transmitting across successive scheduling intervals  till at every user the accumulated information    exceeds some threshold $\Theta$. The threshold rate $\Theta$ is chosen such that enough information has been collected in order to reliably decode the transmitted message, for example via rateless coding/decoding \cite[Ch. 50]{mackay_book}.
  Under this scenario, the incurred delay at the $k$-th user (in terms of the number of scheduling intervals) to decode the transmitted message is given by
\begin{align}
D_k(\{\bbW^{\tau}\}) := \min \left\{t \in \mathbb Z_+ : \textstyle \sum_{\tau=1}^{Lt} R_k^{\tau}(\bbW^\tau) \geq \Theta \right\}. \label{delayk}
\end{align}
 Note that in (\ref{delayk}) we have assumed a quasi-static fading setup, where within the time horizon of interest  the channel per user $k$ remains invariant across all scheduling intervals, i.e., $\bbH_k^{tL+\ell}= \bbH_k^{\ell}$,  $ 1\leq \ell\leq L$ and  $t \in \mathbb Z_+$.
  This assumption is reasonable for instance over  a wideband orthogonal frequency division multiplexing based multiple-access (OFDMA) system, where the users have low mobility. There each scheduling interval comprises of consecutive OFDM symbols and several such scheduling intervals are within the coherence time. Each slot in a scheduling interval is formed by a set of consecutive sub-carriers and OFDM symbols, where the set of consecutive sub-carriers is well within the coherence bandwidth so that each slot can be represented by one channel matrix.
Then, the goal is to jointly design a sequence of precoders $\{\bbW^\tau\}$ which together minimize the weighted sum  delay among all the $K$ users; that is,
\begin{empheq}[box=\widefbox]{align} \label{p2}
(P2)~~~~~\min_{\{\bbW^\tau\in\ccalC\}} ~~ \sum_{k=1}^K ~~ \mu_k D_k(\{\bbW^\tau\})  ~~~~~
\end{empheq}
%
where the weights $\{\mu_k\}_{k=1}^K$  determine each user's priority. Furthermore, to specify the constraints on the precoding matrices for both (P1) and (P2), two interesting codebook scenarios are introduced, as explained below.
\renewcommand{\labelenumi}{\textsf{C\theenumi.}~}
\begin{enumerate}
  \item \textsf{Continuous codebook $\ccalC_c$.}  allows the precoder $\bbW$ to be any arbitrary complex-valued matrix subject to norm and dimensionality constraints. As a result of limited computational capability at the users, the BS can afford to simultaneously transmit at most $d\geq 1$ symbol streams, where we note that a larger $d$ increases the corresponding decoding complexity. Then, incorporating the transmitter power constraint  the continuous codebook can be specified as
      \begin{align}
      \ccalC_c :=\left\{\left.\bbW \in \mathbb C^{M\times d}~ \right|~\|\bbW\|_F^2 \leq P\right\}. \label{Cc}
      \end{align}
Note that the continuous codebook is applicable in a scenario where  over each slot  of every  scheduling interval, pilots precoded by the chosen precoder $\bbW^{\tau}$ can be transmitted so that each user $k$ can directly estimate $\bbH_k^{\tau}\bbW^{\tau}$.

  \item \textsf{Discrete codebook $\ccalC_d$.}
   Such a codebook is motivated by a practical scenario where precoded pilots are not available and where the signaling overhead (needed to indicate the choice of precoder to the users) is limited.  In this case the BS can use a precoder that is formed by concatenating precoders   from a known base codebook $\ccalW$ comprising of a finite number of matrix codewords. It is assumed that $\|\bbW'\|_F^2 =1$, $\forall\; \bbW' \in \ccalW$. Let $\ude = (\bbW',r,p)$ denote an element, where $\bbW'\in \ccalW$, $r$ equals to the column dimension (and rank) of $\bbW'$ such that $\bbW' \in \mathbb C^{M\times r}$, and $p$ determines the power level by which  $\bbW'$ can be scaled.
    Further, let $\udE = \{\ude =(\bbW',r,p): \bbw\in\ccalW, r = \mathrm{rank}(\bbW') \in \mathbb Z_{+} \}$ denote the ground set of all possible such elements, which is known to the BS (and to all users) in advance. For any such element in $\udE$ we adopt the convention that
      \begin{align}
      \ude =(\bbW',r,p) ~\Rightarrow ~\bbW_{\ude} = \bbW'~;~ r_\ude = r~; ~p_{\ude} = p~.\label{econv}
      \end{align}
      Thus, each precoder in $\ccalC_d$   corresponds to some subset of elements $\udU \subseteq \udE$, as given by
      \begin{align}
      \ccalC_d := \left\{\bbW ~\left|~  \exists~ \udU \subseteq \udE~, \bbW :=\left[\{\sqrt{p}_{\ude}\bbW_\ude\}_{\ude \in \udU}\right], r_\udU \leq d,   p_\udU \leq P \right.\right\} \label{Cd}
      \end{align}
      where we follow the notational convention
      \begin{align}
      \udU \subseteq \udE ~\Rightarrow ~ r_\udU = \sum_{\ude\in\udU} r_\ude~; ~p_{\udU} = \sum_{\ude\in\udU}p_\ude~.\label{Uconv}
      \end{align}
      As the counterpart of the matrix dimension constraint in $\ccalC_c$, the sum dimension one of \eqref{Cd} ensures at most $d$ streams are  transmitted. In addition, the sum power constraint of \eqref{Cd} is akin to the Frobenius norm one in \eqref{Cc}.    Note that the concatenation based approach of designing $\ccalC_d$ has a smaller memory footprint, facilitates simpler search algorithms for determining a suitable precoder and can also reduce the signaling burden compared to a finite albeit unstructured codebook.
\end{enumerate}

With these two codebook settings, the goal is to design the precoder matrix (matrices), bearing in mind the aforementioned criteria in (P1) and (P2). The next section will address the first problem of maximizing the instantaneous throughput. In what follows, we collect essential results that follow directly from known results as lemmas (after proper citation) and collect the novel results in propositions.



\section{Maximizing the Instantaneous Throughput}
\label{sec:maxinst}

This section focuses on the one-snapshot problem (P1), which maximizes the minimum among the rates achievable at all the $K$ users for any given time instance. As mentioned earlier, in this whole section the slot index $\tau$ can be dropped for simplicity. 

\subsection{Continuous Codebook}
\label{sec:cont_maxinst}

Notice that the problem (P1) with the continuous codebook $\ccalC_c$ is an NP-hard problem since the particular case with $d=1$ is known to be NP-hard \cite{sidiro06}. Then, to efficiently obtain sub-optimal solutions, it is useful to first consider a simple linear decoding scheme at each user. To this end, denote $\bbG_k \in \mathbb C^{N_k\times d}$ as the linear receive filter per user $k$. With the system model \eqref{kuser_ioc}, the output of the $k$-th receive filter can be expressed as
\begin{align}
\hhatbbs_k = \bbG_k^\dag \bby_k = \bbG_k^\dag \bbH_k \bbW \bbs + \bbG_k^\dag \bbz_k,~~\forall k,\label{kuser_filter}
\end{align}
with the corresponding mean-squared error (MSE) matrix of estimating the signal $\bbs$ given by
\begin{align}
\nonumber\bbE_k(\bbG_k, \bbW) = \mathbb E\left[\left(\hhatbbs_k - \bbs\right)\left(\hhatbbs_k - \bbs\right)^\dag \right]\\
=\left(\bbG_k^\dag \bbH_k \bbW- \bbI_d\right)\left(\bbG_k^\dag \bbH_k \bbW- \bbI_d\right)^\dag + \bbG_k^\dag \bbG_k.\label{kuser_mse}
\end{align}
Interestingly, the MSE matrix $\bbE_k(\bbG_k, \bbW)$ in \eqref{kuser_mse} can be related to the achievable rate $R_k(\bbW)$ of \eqref{p1}, as detailed in the following lemma (cf. \cite{christen}).

\begin{lemma}\label{lem:mse}
For a given precoding matrix $\bbW$, the achievable rate $R_k(\bbW)$ per user $k$ in \eqref{p1} can be obtained by solving the optimal receive filter problem as follows:
\begin{align}
R_k(\bbW) = \max_{\bbG_k} ~\log \left|\bbE_k^{-1} (\bbG_k, \bbW) \right|\label{RkEk}
\end{align}
where its optimum is attained at the linear minimum MSE (LMMSE) filter for the $k$-th user; that is,
\begin{align}
\hhatbbG_k = \left(\bbH_k\bbW\bbW^\dag \bbH_k^\dag + \bbI_{N_k}\right)^{-1}\bbH_k \bbW. \label{optGk}
\end{align}
\end{lemma}
\vspace{-2mm}

Unfortunately, the variables $\{\bbG_k\}$ and $\bbW$ together do not allow decomposing \eqref{p1} into solvable sub-problems.   
Consequently, we introduce more auxiliary variables which allow  us to decompose \eqref{p1} to optimally solvable sub-problems. Towards that end, we state the following lemma which was proposed and used to design precoders over  the MIMO broadcast channel (with unicast transmissions) in \cite{christen} and later for the MIMO interference channel in \cite{jose_nectr10}.

%
%

\begin{lemma}\label{lem:SE}
For any given precoder $\bbW \in \mathbb C^{M\times d}$ and any  filter $\bbG_k \in \mathbb C^{N_k\times d}$, the MSE matrix   $\bbE_k(\bbG_k,\bbW)$ is positive definite and  the following holds
\begin{align}
\max_{\bbS_k \in \mathbb C^{d\times d}: \bbS_k\succ \bb0} ~ \{-\tr(\bbS_k\bbE_k) + \log | \bbS_k|+d \} = \log |\bbE_k(\bbG_k,\bbW)^{-1}|, \label{lemSE}
\end{align}
where the optimum is attained at $\bbS_k = \bbE_k(\bbG_k,\bbW)^{-1}$.
\end{lemma}

It can be verified that for a given precoder $\bbW$ and any given $\bbS_k\succ \bb0$ the solution to
 $\min_{\bbG_k } ~ \tr(\bbS_k\bbE_k(\bbG_k,\bbW))$ is also achieved at (\ref{optGk}).
Then, to   make the problem decomposable, using Lemma \ref{lem:SE} introduce the (matrix) slack variables $\{\bbS_k \in \mathbb C^{d\times d}\}_{k=1}^K$, one per user $k$. With the equivalence asserted in Lemmas \ref{lem:mse} and \ref{lem:SE},  and using the continuous codebook $\ccalC_c$ in \eqref{Cc}, the instantaneous throughput maximization problem (P1) can be reformulated as
\begin{empheq}[box=\widefbox]{align} \label{p1c}
\max_{\|\bbW\|_F^2\leq P\atop \{\bbG_k,\bbS_k\succ \bb0\}} ~~ \min_{k=1,\ldots,K} ~~ -\tr[\bbS_k\bbE_k(\bbG_k,\bbW)] + \log | \bbS_k |+d
\end{empheq}
where the MSE matrix $\bbE_k(\bbG_k,\bbW)$ is given by \eqref{kuser_mse}.

Interestingly, not only the reformulated problem \eqref{p1c} is  equivalent to (P1), each stationary point of \eqref{p1c}  also yields a stationary point of (P1). The latter fact follows upon invoking the gradient expressions given in \cite{christen} and is shown in the sequel.  Further, the reformulated problem \eqref{p1c} also allows us to use cyclic alternating ascent (CAA) algorithm to decompose it into   sub-problems that are solvable. For a fixed $\bbW$, the problem in \eqref{p1c} can be be optimally solved over $\{\bbG_k,\bbS_k\}$. This is  because upon further fixing $\{\bbS_k\succ {\bf 0}\}$, the problem in \eqref{p1c} reduces to that of minimizing the weighted MSE cost over linear filter $\bbG_k$ per user $k$, with the closed-form solution given by \eqref{optGk}; then using those $\{\hat{\bbG}_k\}$, it reduces to the problem in Lemma \ref{lem:SE}, which admits closed-form solution $(\bbE_k(\hat{\bbG}_k,\bbW))^{-1}=\bbW^\dag \bbH_k^\dag\bbH_k\bbW + \bbI_{d}$. A slightly more complicated sub-problem appears when solving the precoder $\bbW$ while fixing both $\{\bbG_k\}$ and $\{\bbS_k\}$. To tackle this sub-problem, consider its equivalent form given by
\begin{subequations}\label{p1cbeta}
\begin{align}
\max_{\|\bbW\|_F^2\leq P} ~~& ~~~~~\beta \label{p1cbeta1}\\
\mathrm{s. to} ~~&~ -\tr[\bbS_k\bbE_k(\bbG_k,\bbW)] + \log | \bbS_k |+d \geq \beta, ~~\forall k \label{p1cbeta2}
\end{align}
\end{subequations}
where at the optimum of \eqref{p1cbeta}, $\beta$ becomes equal to the minimum of achievable costs among all $K$ users. Furthermore, define the following Cholesky factorization per user $k$ as $\bbS_k = \bbB_k \bbB_k^\dag$, and  thus the constraint \eqref{p1cbeta2} can be cast as a quadratic cone one in terms of the variable $\bbW$, and the sub-problem for solving it becomes
\begin{subequations}\label{p1cW}
\begin{align}
\max_{\|\bbW\|_F^2\leq P} ~~& ~~~~~\beta \label{p1cW1}\\
\mathrm{s. to} ~~&~ c_k -\beta \geq \left\|\bbB_k ^\dag \left( \bbG_k^\dag \bbH_k \bbW- \bbI_d \right) \right\|_F^2, ~~\forall k \label{p1cW2}
\end{align}
\end{subequations}
where the constant $c_k :=  \log | \bbS_k |+d- \left\|\bbG_k \bbB_k \right\|_F^2$. Notice that the other power constraint on $\bbW$ is a quadratic one, hence, the sub-problem \eqref{p1cW} for obtaining $\bbW$ while fixing the others is a second-order cone program (SOCP), and thus can be solved efficiently using some off-the-shelf optimization tools, e.g., the interior point optimization routine in SeDuMi
\cite{sedumi}.

These aforementioned sub-problems suggest an iterative CAA algorithm yielding
successive estimates of one of the two groups of variables -- $\{\bbG_k,\bbS_k\}$, and $\bbW$ -- with the remaining group fixed, as tabulated in Algorithm \ref{alg:p1ad}. The convergence of Algorithm \ref{alg:p1ad}  in terms of the  objective value is guaranteed due to the cyclic ascent nature of the algorithm that ensures a monotonically non-decreasing objective across iterations. However, proving the convergence for the sequence of iterates is more involved. The following convergence claim applies for  Algorithm \ref{alg:p1ad}  when it is invoked without any limit on the number of iterations. A similar CAA convergence result is outlined in \cite{yafeng}, but for a different problem setup involving MIMO interference channels.

\begin{algorithm}[htb]
\caption{: (P1) with $\ccalC_c$. Input the channel matrices $\{\bbH_k\}_{k=1}^K$, and an initial feasible $\bbW$. Output the iterates upon convergence.} 
\begin{algorithmic}[1] \label{alg:p1ad}

\WHILE{the iterates converge or maximum number of iterations is reached}
    \FOR{$k = 1, \ldots, K$}
        \STATE Obtain the  LMMSE optimal receive filter as $\bbG_k \leftarrow \left(\bbH_k\bbW\bbW^\dag \bbH_k^\dag + \bbI_{N_k}\right)^{-1} \bbH_k \bbW$ .
        \STATE Update the slack matrix $\bbS_k \leftarrow \bbW^\dag \bbH_k^\dag\bbH_k\bbW + \bbI_{d}$, with the MSE matrix calculated via \eqref{kuser_mse}.
    \ENDFOR
    \STATE Obtain the precoder matrix $\bbW$ by solving the SOCP problem \eqref{p1cW}.
\ENDWHILE
\end{algorithmic}
\end{algorithm}

\begin{proposition}\label{prop:p1adconvg}
Either the sequence of iterates generated by Algorithm \ref{alg:p1ad} converges to a stationary point or each of its accumulation points is a stationary point of (P1), and the objective is  non-decreasing as the iterations proceed.
\end{proposition}

\proof The proof is given in Appendix \ref{app:cca}.

\subsection{Discrete Codebook}
\label{sec:dist_maxinst}

From the definition of $\ccalC_d$ in \eqref{Cd}, each valid precoder corresponds to a subset $\udU \subseteq \udE$. Thus, the achievable rate for user $k$ in \eqref{Rkt} can be considered as a set function $R_k: 2^\udE \rightarrow \mathbb R_{+}$ given by
\begin{align}
R_k(\udU) = \log \left|\bbI + \sum_{\ude\in\udU} p_\ude \bbH_k \bbW_\ude \bbW_\ude^\dag \bbH_k^\dag\right|, ~~\forall k \label{Rktset}
\end{align}
for all $\udU \subseteq \udE$.
We offer the following useful result.

%
\begin{proposition}\label{prop:submod1}
The set function $R_k(\cdot)$ in \eqref{Rktset} is a submodular set function, i.e.,
\begin{align}
R_k(\udU\cup\{\ude\})  - R_k(\udU) \geq R_k(\udU'\cup\{\ude\})  - R_k(\udU'), ~~\forall k, \label{Rktsub}
\end{align}
for all $\udU \subseteq \udU' \subseteq \udE$ and $\ude \in \udE$. Further, it is also monotonic as $R_k(\udU) \leq R_k(\udU')$, $\forall \udU \subseteq \udU'$, and normalized $R_k(\emptyset) = 0$ where $\emptyset$ denotes the empty set.
\end{proposition}
\proof The proof is given in Appendix \ref{app:subm}.

Thus, (P1) with the discrete codebook $\ccalC_d$ in \eqref{Cd} becomes a robust submodular function maximization problem,   given by
%
\begin{empheq}[box=\widefbox]{align}
 \max_{\udU\subseteq\udE} &~~ \min_{k=1,\ldots,K} ~~ R_k(\udU) ~~\mathrm{s. to}~~ r_\udU \leq d, ~  p_\udU \leq P\label{p1d}
\end{empheq}
%
For general submodular set functions, maximizing a robust criterion with even one  constraint has been shown to be strongly NP hard \cite{krause_jmlr08}. Here, we show that for the particular submodular set functions given in \eqref{Rktset},  the robust rate maximization problem in (\ref{p1d}) with   only the power constraint, i.e., the problem
\begin{empheq}[box=\widefbox]{align}
\max_{\udU\subseteq\udE} ~~ \min_{k=1,\ldots,K} ~~ R_k(\udU) ~~
\mathrm{s. to}~~    p_\udU \leq P\label{p1dss}
\end{empheq}
is also strongly NP hard, as asserted in Proposition \ref{prop:p1dnp}. 
Note that   an instance of the problem in (\ref{p1dss}) comprises of: the number of users $K$ along with their channel matrices $\{\bbH_k\}_{k=1}^K$, the set $\udE$ (specified via a base code book $\ccalW$ of precoders and a power level for each precoder in $\ccalW$) as well as the power budget $P$.
In particular, we show that \eqref{p1dss} is NP hard even over instances where we restrict $K=O(|\udE|^{\Delta})$ for any arbitrarily fixed positive integer $\Delta\geq 2$.

\begin{proposition}\label{prop:p1dnp}
Unless P=NP, there cannot exist any polynomial time approximation algorithm for \eqref{p1dss}. More precisely: If there exists a positive function $\gamma: \mathbb Z_{+}\rightarrow \mathbb R_{+}$ and an algorithm that, for all $|\udE|$ and $P$, in time polynomial in $|\udE|$, is guaranteed to find a subset $\udU'$ satisfying the power constraint $P$ such that
\begin{align}
\min_{k} R_k(\udU') \geq \gamma(|\udE|) \max_{\sum_{\udU\subseteq\udE} p_\udU \leq P} \min_k R_k(\udU),
\end{align}
then P=NP.
\end{proposition}
\proof The proof is given in Appendix \ref{app:np}.

Proposition \ref{prop:p1dnp}   manifests the hardness of the discrete precoder design problem, and that polynomial-complexity algorithms cannot approximate the optimal rate within a bound that is only determined by   $|\udE|$. In the following, we consider the problem \eqref{p1dss} and adopt a bicriterion optimization approach. We leverage the \emph{Submodular Saturation algorithm} (SSA) developed in \cite{krause_jmlr08}, which considers the general robust submodular minimization problem but can offer guarantees only for integral valued submodular functions. Since the submodular functions that are of interest to us are not integral valued, we modify the SSA by using recent results for the submodular set-cover problem, wherein the submodular cost function can be real-valued \cite{goyal_inf}.

Following the SSA, the proposed algorithm exploits the idea of the bisection method   which is applied to the following  equivalent formulation of \eqref{p1dss}:
\begin{align}\label{p1de}
\{\hhatc, \hat{\udU}\}:= \arg \max_{c,\udU~\subseteq\udE} &~~  c,~~~~~
\mathrm{s. to}~~  R_k(\udU) \geq c,~\forall k \; \mathrm{and}\;p_\udU \leq P.
\end{align}
The equivalence between \eqref{p1de} and \eqref{p1dss} holds, since at the optimum of \eqref{p1de}, the value $\hhatc$ will always be equal to the minimum of $\{R_k(\hat{\udU})\}$ across all the $K$ users. Now suppose that there exists an algorithm that, for any given value $c$, solves the following optimization problem
\begin{align}\label{p1dbi}
\hat{\udU}_c:=\arg \min_{\udU~\subseteq\udE} &~~ p_\udU,~~~~~
\mathrm{s. to}~~  R_k(\udU) \geq c,~\forall k=1,\ldots,K,
\end{align}
then the power associated with the optimum set $\hat{\udU}_c$ can be used to decide the relationship between the prescribed value $c$ and the optimum $\hhatc$ in \eqref{p1de}. Specifically, if it turns out that $p_{\hat{\udU}_c} \leq P$, then $c$ is feasible for \eqref{p1de} and it must hold that $c \leq \hhatc$. Otherwise, the chosen value $c$ is infeasible for \eqref{p1de} and we have $c > \hhatc$. Hence, an iterative binary search on $c$ would then allow us to find the maximum value that is feasible to \eqref{p1de}. However, the problem \eqref{p1dbi} is not exactly solvable, but can only be approximated as shown below.

To illustrate this, consider any feasible value $c$ and the truncated function $R_{k,c}(\udU):= \min\{R_k(\udU),c\}$. Let $\bbarR_c(\udU) := (1/K) \sum_{k=1}^K R_{k,c}(\udU)$ be their average function, which is also submodular and monotonic (follows from a result in \cite{krause_jmlr08}). With these definitions, we have $\bbarR_c(\udE) = c$, and the constraint in \eqref{p1dbi} holds if and only if $\bbarR_c(\udU) = c$, which establishes the equivalence between \eqref{p1dbi} and the following one
\begin{align}\label{p1dbi2}
\hat{\udU}_c:=\arg \min_{\udU~\subseteq\udE} &~~ p_\udU,~~~~~
\mathrm{s. to}~~  \bbarR_c(\udU) = \bbarR_c(\udE).
\end{align}
Interestingly, the reformulated problem \eqref{p1dbi2} is an instance of the \emph{submodular covering problems}. A greedy algorithm has been proposed in  \cite{wolsey_comb82} to approximately solve such problems but that algorithm yields a useful guarantee only for integral valued submodular functions.
Recall that the submodular functions $\{R_k(.)\}$ in \eqref{Rktset} are not integral-valued. Consequently, we employ a variation of the greedy algorithm proposed in \cite{goyal_inf} and given here in Algorithm \ref{alg:gpcM}.

\begin{algorithm}[htb]
\caption{: \eqref{p1dbi2} with a feasible $c$. Input the channel matrices $\{\bbH_k\}$,   $\delta\in(0,1)$ and   the ground set $\udE$. Output the greedy solution $\hat{\udU}_G$ to \eqref{p1dbi2}.} 
\begin{algorithmic} \label{alg:gpcM}
\STATE Initialize $\hat{\udU}_G = \emptyset$.
\WHILE{$\bbarR_c(\hat{\udU}_G) < c(1-\delta)$}
    \STATE Update $\hat{\udU}_G \leftarrow$
    $\hat{\udU}_G \cup \left\{ \arg \max_{\ude \in \udE \setminus \hat{\udU}_G} \frac{\bbarR_c(\hat{\udU}_G \cup\{\ude\})-\bbarR_c(\hat{\udU}_G)}{p_\ude} \right\}.$
\ENDWHILE
\end{algorithmic}
\end{algorithm}

The following lemma follows from Theorem 1 of \cite{goyal_inf} when the latter is invoked using submodular set function $\bbarR_c(.)$, threshold $c$ (which we note is feasible for $\bbarR_c(.)$, i.e., $\bbarR_c(\udE)\geq c$) and a gap $c\delta$, where $\delta\in(0,1)$.

\begin{lemma}\label{lem:gpcM}
With a monotonic real-valued submodular function  $\bbarR_c(.)$, any $\delta\in(0,1)$  and a (feasible) value $c$, Algorithm \ref{alg:gpcM} finds a set $\hat{\udU}_G$ such that $\bbarR_c(\hat{\udU}_G) \geq c(1-\delta)$ and $p_{\hat{\udU}_G} \leq  p_{\hat{\udU}_c}(1+\ln(1/\delta))$, where $\hat{\udU}_c$ is an optimal solution to \eqref{p1dbi2}.
\end{lemma}

Note that the greedy   Algorithm \ref{alg:gpcM} can only approximate the optimal solution $\hat{\udU}_c$. This prevents from implementing the bisection method based on the equivalence between \eqref{p1dss} and \eqref{p1de}, since solving the latter requires to  find the exact optimal solution $\hat{\udU}_c$ to \eqref{p1dbi} per bisection iteration for any given $c$. Therefore, we need to adapt the original binary search procedure in order to accommodate the greedy approximation algorithm. In particular, for any   specified $\delta\in(0,1)$, the binary search criteria budget per iteration is scaled to $P(1+\ln(1/\delta))$, and the corresponding decision rule is also changed as follows: if Algorithm \ref{alg:gpcM} outputs $p_{\hat{\udU}_c} >  P(1+\ln(1/\delta))$, the  chosen value $c$ is infeasible to \eqref{p1de} and $c > \hhatc$; otherwise, the output $\hat{\udU}_c$ is a feasible solution to a relaxed version of \eqref{p1de} with   budget  $P(1+\ln(1/\delta))$, and will be kept as the best current solution to it.  Such adapted bisection method is tabulated in Algorithm \ref{alg:bis} which is polynomial time (for any fixed $\epsilon$) and  has the following optimality, as asserted in the following proposition.

\begin{algorithm}[htb]
\caption{: \eqref{p1de} with $\ccalC_d$. Input the channel matrices $\{\bbH_k\}$, $\delta$, tolerance $\epsilon$ and the ground set $\udE$. Output the set $\hat{\udU}$.} 
\begin{algorithmic} \label{alg:bis}
\STATE Initialize $c_{min} = 0$, $c_{max} = \min_k R_k(\udE)$, and $\hat{\udU} = \emptyset$.
\WHILE{$c_{max} - c_{min} > \epsilon$}
    \STATE Set $c \leftarrow (c_{min}+c_{max})/2$, and define $\bbarR_c(\udU) := (1/K) \sum_{k=1}^K \min\{R_{k}(\udU),c\}$.
    \STATE Use Algorithm \ref{alg:gpcM} with input $\delta$ to obtain the greedy solution $\hat{\udU}_G$.
    \IF{$p_{\hat{\udU}_G}> P(1+\ln(1/\delta))$}
        \STATE Update $c_{max} \leftarrow c$.
    \ELSE
        \STATE Update $c_{min} \leftarrow c$ and $\hat{\udU} \leftarrow \hat{\udU}_G$.
    \ENDIF
\ENDWHILE
\end{algorithmic}
\end{algorithm}

\begin{proposition}\label{prop:p2bis}
For any power budget $P$  and given $\delta,\epsilon\in(0,1)$, Algorithm \ref{alg:bis} finds a solution $\hat{\udU}$ such that
\begin{align}
\min_{k} R_k(\hat{\udU}) \geq (1-K\delta)\max_{\udU:p_\udU \leq P} \min_{k} R_k(\udU)-\epsilon(1-K\delta)
\end{align}
 and $p_{\hat{\udU}} \leq  P(1+\ln(1/\delta))$.
\end{proposition}
\proof The proof is given in Appendix  \ref{app:alg4}.

\noindent {\bf Remark 1:} In practice Algorithm \ref{alg:bis} gives  good results when invoked with $\delta=0$ but where the condition $p_{\hat{\udU}_G}> P(1+\ln(1/\delta))$ is replaced by $p_{\hat{\udU}_G}> P$. In addition, simple enhancements such as replacing the search space $\ude \in \udE \setminus \hat{\udU}_G$ in Algorithm \ref{alg:gpcM} with $\ude \in \udE \setminus \hat{\udU}_G:p_{\ude \cup \hat{\udU}_G}\leq P$ (when the latter is invoked by Algorithm \ref{alg:bis}) also improve performance.

Finally, it is useful to derive an upper bound for \eqref{p1dss} to benchmark the performance of Algorithm \ref{alg:bis},  as given by
\begin{align}\label{eq:UBmaxmin}
\nonumber\max_{\{x_{\ude}\}_{\ude\in\udE},\beta}& ~~\beta\\
\nonumber\log \left|\bbI +  \sum_{\ude\in\udE}p_{\ude}x_{\ude}\bbH_k  \bbW_{\ude}\bbW^{\dag}_{\ude} \bbH_k^{ \dag}\right|&\geq\beta,\;\forall\;k,\\
\sum_{\ude\in\udE}x_{\ude}p_{\ude}\leq P 0\leq x_{\ude}&\leq 1,\;\forall\;\ude\in\udE .
\end{align}
Notice that (\ref{p1dss}) and (\ref{eq:UBmaxmin}) are equivalent if we enforce stricter constraints $x_{\ude}\in\{0,1\},\;\forall\;\ude\in\udE$ in (\ref{eq:UBmaxmin}). Then, an important observation that can be made using \cite{Boyd_convex} pp. 74, is that for each $1\leq k\leq K$, the function $\log \left|\bbI +  \sum_{\ude\in\udE}p_{\ude}x_{\ude}\bbH_k  \bbW_{\ude}\bbW^{\dag}_{\ude} \bbH_k^{ \dag}\right|$ is jointly concave in $[p_{\ude}]_{\ude\in\udE}\in\mathbb R_{+}^{|\udE|}$. Consequently, it follows that (\ref{eq:UBmaxmin}) is a convex optimization problem that can be efficiently solved.

%


\section{Minimizing the Weighted Sum Delay}
\label{sec:mindelay}


A different precoder design criterion is considered in this section. Specifically, in contrast to focusing on the minimum instantaneous throughput among the users as in Section \ref{sec:maxinst}, a weighted performance in terms of decoding delay across the $K$ users  (P2) becomes the subject of interest. In order to make (P2) more tractable, first consider a different expression for the decoding delay   given by
\begin{align}
D_k(\{\bbW^\tau\}) := 1 + \sum_{t=1}^{\infty} \left[1- \ind\left( \textstyle \sum_{\tau=1}^{Lt} R_k^\tau(\bbW^\tau)/\Theta \right) \right], ~~\forall k \label{delayk2}
\end{align}
where the function $\ind(\cdot)$   defined as $\ind(\chi) = 1$ if $\chi \geq 1$ and 0 if $0\leq \chi< 1$, indicates whether the threshold $\Theta$ has been reached for the accumulated rate per user $k$. (Notice that the function value for negative $\chi$ can be disregarded, since the achievable rate is always non-negative.) Clearly, in \eqref{delayk2} the delay $D_k$ simply counts the number of scheduling intervals that are needed for the non-decreasing accumulated rate to cross $\Theta$ and  the common message to be reliably decoded. Although the delay $D_k$ can be expressed as an analytical function of the precoder matrices as in \eqref{delayk2},   the indicator function $\ind(\cdot)$ still makes the problem (P2) difficult to solve.
In the following, we will first consider optimizing (P2) over the continuous codebook and then the optimization over the discrete codebook.

\subsection{Continuous Codebook}
\label{sec:cont_mindelay}

We consider solving (P2) with the continuous codebook $\{\ccalC_c\}$. Notice that the indicator function $\ind(\chi)$ is discontinuous at the point $\chi=1$ and this discontinuity in the cost as a function of the accumulated rate will render it difficult to optimize the precoding codewords. Even upon employing an alternating optimization approach as in Section \ref{sec:cont_maxinst}, the resultant sub-problems are non-convex and not easily solvable.
As the difficulty lies in the discontinuity,  we propose to relax the indicator function as
\begin{align}
\ind_r(\chi) = \left\{\begin{array}{cl}
                   \chi & \mathrm{if~} 0\leq \chi \leq 1 \\
                   1 & \mathrm{otherwise}
                 \end{array}
 \right. =  
                   \min \{ \chi,1\} & \;\forall\; \chi \geq 0. 
\label{indrel}
\end{align}
Since the accumulated rate is never negative, the weighted sum delay minimization problem (P2), with the delay $D_k(\{\bbW^\tau\})$  lower bounded by substituting the relaxed $\ind_r(\cdot)$ of \eqref{indrel} into \eqref{delayk2}, is relaxed to
\begin{empheq}[box=\widefbox]{align} \label{p2rel}
(P2'):~~\sum_{k=1}^K\mu_k + ~~~\min_{\{\bbW^\tau\in\ccalC_c\}} ~~ \sum_{k=1}^K ~\sum_{t=1}^{\infty}\mu_k\left[ 1-  \min \left\{\textstyle \sum_{\tau=1}^{Lt} R_k^\tau(\bbW^\tau)/\Theta, 1\right\}\right]. ~~~~~
\end{empheq}
%


We offer the following result.

\begin{proposition}\label{prop:propoFK}
Suppose that for each user $k:1\leq k\leq K$ the input channel set $\{\bbH^\tau_k\}_{\tau=1}^L$  has at-least one non-zero channel matrix  and further suppose that  $\Theta$ is finite. Then, $\exists\;\hat{t}<\infty$ such that any optimal solution $\{\bbW_{\rm opt}^\tau\}$ to (P2$'$) can be truncated by setting $ \bbW_{\rm opt}^\tau={\bf 0}$ for all $\tau>L\hat{t}$, without sacrificing optimality. \end{proposition}
\proof The proof is given in Appendix   \ref{app:propoFK}.

We emphasize that $\hat{t}$ in Proposition \ref{prop:propoFK} can be determined as a function of only the given input channel set and the threhold. Clearly the truncation can be done without loss of optimality by  by setting $\bbW_{\rm opt}^\tau={\bf 0}$ for all $\tau>Lt'$, for any $t'\geq \hat{t}$, as well.
Further, note that   Proposition \ref{prop:propoFK} implies that without loss of optimality (P2$'$) can be regarded as a finite dimensional optimization problem in which the set of feasible solutions is compact.
Next, in order to solve the relaxed problem (P2$'$), we adopt the following approach. We start by considering a particular choice $\tilde{t}$ for the number of scheduling intervals 
and pose the following problem
\begin{empheq}[box=\widefbox]{align} \label{p2rel}
(P2'')~~\max_{\{\bbW^\tau\in\ccalC_c\}} ~~ \sum_{k=1}^K ~\sum_{t=1}^{\tilde{t}}\mu_k  \min \left\{\textstyle \sum_{\tau=1}^{Lt} R_k^\tau(\bbW^\tau)/\Theta, 1\right\}. ~~~~~
\end{empheq}
Then, to solve (P2$''$) we leverage
the same approach as in Section \ref{sec:cont_maxinst} and decompose it into optimally solvable sub-problems. To this end, we introduce the linear filters $\{\bbG_k^\tau \in \mathbb C^{N_k\times d}\}_{\tau=1}^{L\tilde{t}}$ with corresponding MSE matrices $\{\bbE_k^\tau(\bbG_k^\tau, \bbW^\tau)\}_{\tau=1}^{L\tilde{t}}$ as in \eqref{kuser_mse} and the matrix slack variables $\{\bbS_k^\tau \in \mathbb C^{d\times d}\}_{\tau=1}^{L\tilde{t}}$, per user $k$. Invoking Lemmas \ref{lem:mse} and \ref{lem:SE},  the   problem (P2$''$) can be written as
\begin{empheq}[box=\widefbox]{align} \label{p2rel2}
\max_{\{\bbW^\tau\in\mathbb C^{M\times d}:\|\bbW^\tau\|_F^2\leq P\}, \{\bbG_k^\tau\},\{\bbS_k^\tau\succ\bb0\}}  \sum_{k=1}^K \sum_{t=1}^{\tilde{t}} \mu_k \min \left\{ \frac{1}{\Theta} \sum_{\tau=1}^{Lt} \left[-\tr\left[\bbS_k^\tau ~\bbE_k^\tau(\bbG_k^\tau, \bbW^\tau)\right]+\log |\bbS_k^\tau| +d \right], 1 \right\}.
\end{empheq}

Hence, the CAA algorithm is applicable to the relaxed problem (\ref{p2rel2}). Fixing $\{\bbW^\tau\}$, the problem in (\ref{p2rel2}) can be optimally solved over $\{\bbG_k^\tau,\bbS_k^\tau\}$, using Lemmas \ref{lem:mse} and \ref{lem:SE}. It now remains to update all the precoders $\{\bbW^\tau\}$, while fixing the other variables, $\{\bbG_k^\tau\}$ and $\{\bbS_k^\tau\}$. Using $\{\alpha_k^t\}_{t=1}^{\tilde{t}}$ to denote the minimum between the accumulated normalized rate and the unit threshold for each user $k$, the aforementioned sub-problem for $\{\bbW^\tau\}$ is equivalent to
\begin{subequations}\label{p2calpha}
\begin{align}
\max_{\{\bbW^\tau\in\mathbb C^{M\times d}:\|\bbW^\tau\|_F^2\leq P\},\{\alpha_k^t\}} ~~& ~~~\sum_{k=1}^K \sum_{t=1}^{\tilde{t}}~  \mu_k \alpha_k^t  \label{p2calpha1}\\
\mathrm{s. to} ~~&~~~ \Theta \alpha_k^t \leq \sum_{\tau=1}^{Lt} \left\{-\tr\left[\bbS_k^\tau ~\bbE_k^\tau(\bbG_k^\tau, \bbW^\tau)\right]+\log |\bbS_k^\tau| +d \right\}  \label{p2calpha2}\\
&~~~ ~~ \alpha_k^t \leq 1  , ~~\forall k, t \label{p2calpha3}
\end{align}
\end{subequations}
Furthermore, defining the Cholesky factorization per user $k$ and slot $\tau$ as $\bbS_k^\tau = \bbB_k^\tau (\bbB_k^\tau)^\dag$,  the problem \eqref{p2calpha} can be reformulated as
\begin{subequations}\label{p2calphab}
\begin{align}
\max_{\{\bbW^\tau\in\mathbb C^{M\times d}:\|\bbW^\tau\|_F^2\leq P\},\{\alpha_k^t,\beta_k^\tau\}} ~~ ~~~\sum_{k=1}^K &\sum_{t=1}^{\tilde{t}}~  \mu_k \alpha_k^t \label{p2calphab1}\\
\mathrm{s. to} ~~~~~ \Theta \alpha_k^t &\leq \sum_{\tau=1}^{Lt} \left\{- \|\bbG_k^\tau \bbB_k^\tau\|_F^2 -\beta_k^\tau +\log |\bbS_k^\tau| +d \right\}  \label{p2calphab2}\\
~~~ \beta_k^\tau & \geq \left\|(\bbB_k^\tau) ^\dag \left[ (\bbG_k^\tau)^\dag \bbH_k^\tau \bbW^\tau- \bbI_d \right] \right\|_F^2, ~~\forall k, \tau\label{p2calphab3}\\
~~~  \alpha_k^t &\leq 1  , ~~\forall k, t. \label{p2calphab4}
\end{align}
\end{subequations}
Thus, the constraint \eqref{p2calpha2} is represented by the linear constraint in \eqref{p2calphab2}  together with a series of quadratic cone constraints in \eqref{p2calphab3}.
Notice that other constraints on the power of each $\bbW^\tau$ are also quadratic. Hence, the sub-problem \eqref{p2calphab} for obtaining $\{\bbW^\tau\}$ while fixing the remaining variables is also an SOCP, and can be solved efficiently as mentioned earlier.

These aforementioned sub-problems with their optimal solutions suggest an iterative CAA algorithm to solve (P2$''$). However, upon convergence the accumulated rates of some users could be below $\Theta$ in which case we can increment $\tilde{t}$ and repeat the process. This procedure is tabulated in Algorithm \ref{alg:p2ad}.
Notice we have assumed that a set of precoders $\{\breve{\bbW}^{(\ell)}\in\ccalC_c\}_{\ell=1}^L$ yielding a rate vector ${\bf\Delta}\succ{\bf 0}$ (componentwise strictly greater than zero) over any scheduling interval is provided as an input. Such a set can be found by using the CAA algorithm of Section   \ref{sec:cont_maxinst} on the input channel matrices
$\{\bbH_k^\ell\},\;1\leq\ell\leq L,1\leq k\leq K$. Indeed, an admission control module can be implemented in which the group of users to receive a common message is decided by verifying whether the  instantaneous rate optimizing algorithm of Section   \ref{sec:cont_maxinst} when used over that group can achieve a strictly positive (or a large enough) value for the minimum instantaneous rate.


\begin{algorithm}[htb]
\caption{: To approximately solve (P2$'$). Input $\tilde{t}$, the channel matrices $\{\bbH_k^\tau\},\;1\leq\tau\leq L,1\leq k\leq K$, a feasible set of precoders $\{\breve{\bbW}^{(\ell)}\}_{\ell=1}^L$ yielding rate vector ${\bf\Delta}\succ{\bf 0}$. Output the final iterates.} 
\begin{algorithmic} \label{alg:p2ad}
\WHILE{For at-least one user $k$ the accumulated rate is below $\Theta-\Delta_k$}
\STATE Increment $\tilde{t}\leftarrow \tilde{t} +1$ and initialize $\{\bbW^\tau\}_{\tau=1}^{L\tilde{t}}$
\REPEAT
    \FOR{$k = 1, \ldots, K$}
    \FOR{$\tau=1,\ldots,L\tilde{t}$}
        \STATE Given the precoder $\bbW^\tau$ at slot $\tau$, update the  LMMSE optimal receive filter as $\bbG_k^\tau \leftarrow \left[\bbH_k^\tau\bbW^\tau(\bbW^\tau)^\dag (\bbH_k^\tau)^\dag + \bbI_{N_k}\right]^{-1} \bbH_k^\tau \bbW^\tau$ .
        \STATE Update the slack matrix $\bbS_k^\tau \leftarrow(\bbW^\tau)^\dag (\bbH_k^\tau)^\dag\bbH_k^\tau\bbW^\tau + \bbI_{d}$. 
    \ENDFOR
    \ENDFOR
    \STATE Obtain the precoder matrices $\{\bbW^\tau\}$ by solving the SOCP problem \eqref{p2calphab}.
\UNTIL{Convergence}
\ENDWHILE
\STATE Let $\{\bbW^\tau\}_{\tau=1}^{L\tilde{t}}$ be the iterate upon convergence or an accumulation point. Output it if accumulated rate of each user is no less than $\Theta$, else augment $\{\bbW^\tau\}_{\tau=1}^{L\tilde{t}}$ by $\{\bbW^{L\tilde{t}+\ell}=\breve{\bbW}^{(\ell)}\}_{\ell=1}^L$ and output it.
\end{algorithmic}
\end{algorithm}


Suppose that a simple initialization, that comprises repeating $\{\breve{\bbW}^{(\ell)}\}_{\ell=1}^L$ over $\tilde{t}$ scheduling intervals, is employed in each outer iteration of Algorithm \ref{alg:p2ad}. We can then prove that the algorithm terminates in a finite number steps even for this simple initialization by employing arguments similar to those in the proof of Proposition \ref{prop:propoFK}, along with the fact that the alternating optimization procedure employed by the algorithm to sub-optimally solve (P2$''$) monotonically improves the objective function value. A stronger result is stated in the following which holds for any feasible initialization. 

\begin{proposition}\label{prop:p2adconvg}
 The output upon termination of Algorithm \ref{alg:p2ad} is a stationary point of (P2$'$).
\end{proposition}
\proof The proof is given in Appendix   \ref{app:p2adconvg}.

Notice that each successive outer iteration of Algorithm \ref{alg:p2ad} involves optimizing over a larger number of variables since the number of scheduling intervals is incremented by one. One variation that can substantially reduce complexity is to only optimize the transmit precoders $\{\bbW^\tau\}$ for $L(\tilde{t}-1)+1\leq\tau\leq L\tilde{t}$ (which correspond to the last scheduling interval) in each outer iteration, and fix the other precoders to their respective values obtained in the previous iterations. It can be proved that this variation also terminates in a finite number of iterations but its output need not be a stationary point of (P2$'$).
\subsection{Discrete Codebook}
\label{sec:dist_mindelay}
In this section, we   consider the discrete codebook version of (P2) given by
\begin{empheq}[box=\widefbox]{align} \label{p2D}
(P2D)~~~~~\min_{\{\bbW^\tau\in\ccalC_d,\;\forall\;\tau\}} ~~ \sum_{k=1}^K ~~ \mu_k D_k(\{\bbW^\tau\})  ~~~~~
\end{empheq}
We assume that a set $\{\breve{\bbW}^{(\ell)}\in\ccalC_d\}_{\ell=1}^L$ is available which achieves a rate-vector ${\bf \Delta}\succ{\bf 0}$ over any scheduling interval.
We will show that (P2D) can be reformulated and sub-optimally solved by using existing algorithms from \cite{gamzu_thesis10,gamzu_soda11} but at a high complexity. We first propose a novel and non-trivial modification to an algorithm from \cite{gamzu_soda11}, which can significantly reduce the complexity  and also offer  a performance guarantee. This modified algorithm is presented here as Algorithm \ref{alg:p2DCB}.
  Notice that Algorithm \ref{alg:p2DCB} involves maintaining an ordered stack $\ccalS$ to which a set of codewords is added in each iteration. Upon termination, the set first added to $\ccalS$ is used in the first scheduling interval, the set added second is used in the second interval and so on. Further, notice that each iteration of Algorithm \ref{alg:p2DCB} also involves (approximately) solving a maximization  problem (\ref{p2ppAlgo}) by invoking Algorithm \ref{alg:p2greedy}.

\begin{algorithm}[htb]
\caption{: To approximately solve (P2D). Input the channel matrices $\{\bbH_k^\tau\},\;1\leq\tau\leq L,1\leq k\leq K$, a feasible set of precoders $\{\breve{\bbW}^{(\ell)}\}_{\ell=1}^L$ yielding rate vector ${\bf\Delta}\succ{\bf 0}$. Output the final iterates.}
\begin{algorithmic} \label{alg:p2DCB}
\STATE Set stack $\ccalS=\phi$, $t=1$,$\ccalI=\{1,\cdots,K\}$ and $\theta_k=0,\;\forall\;k$.
\REPEAT
\STATE Using Algorithm \ref{alg:p2greedy} sub-optimally solve:
 \begin{subequations}
\begin{align}
\max_{\{\bbW^{(\ell)}\in\ccalC_d\}_{\ell=1}^L}\left\{\sum_{k\in\ccalI}\mu_k\min\left\{1,\frac{\sum_{\ell=1}^L R_k^\ell(\bbW^{(\ell)})}{\Theta(1-\theta_k)}\right\}\right\}
  \label{p2ppAlgo}
\end{align}
\end{subequations}
  and obtain $\{\tilde{\bbW}^{(\ell)}\}_{\ell=1}^L$.
\STATE Augment stack $\ccalS$ by adding $\{\bbW^{(t-1)L+\ell}=\tilde{\bbW}^{(\ell)}\}_{\ell=1}^L$ to it, update $\theta_k=\theta_k + \frac{\sum_{\ell=1}^L R_k^\ell(\tilde{\bbW}^{(\ell)})}{\Theta},\;\forall\;k$, $t\leftarrow t+1$ and $\ccalI\leftarrow\ccalI\setminus\{k\in\{1,\cdots,K\}:\theta_k\geq 1\}$.
\UNTIL{$\ccalI=\phi$ or $\theta_k\geq 1 - \Delta_k,\;\forall\;k$}
\STATE Output $\ccalS$ if accumulated rate of each user is no less than $\Theta$, else augment $\ccalS$ by $\{\bbW^{(t-1)L+\ell}=\breve{\bbW}^{(\ell)}\}_{\ell=1}^L$ and output it.
\end{algorithmic}
\end{algorithm}

The submodular property of the   rate functions is utilized again to sub-optimally solve (\ref{p2ppAlgo}) in Algorithm \ref{alg:p2greedy}. We next explain how this algorithm was obtained and then state its performance guarantee.
Since  in (\ref{p2ppAlgo}) the precoders $\{\bbW^{(\ell)}\}_{\ell=1}^L$ across all $L$ slots are design variables, it is necessary to define a concatenated ground set, $\udF$, as
\begin{align}
\udF := \left\{(\ude, \ell) | ~\ude \in \udE \;\&\; \ell\in\{1,\cdots, L\}\right\}.
\end{align}
Then, for any given subset $\ccalI\subseteq\{1,\cdots,K\}$ and any scalars $\theta_k\in[0,1)\;\forall\;k\in\ccalI$, we
define the set function $f: 2^\udF \rightarrow \mathbb R_{+}$, as
\begin{align}
f(\udV) =   \sum_{k\in\ccalI}~ \mu_k \min \left\{ \frac{1}{\Theta(1-\theta_k)} \sum_{\ell=1}^L\log \left|\bbI + \sum_{(\ude,\ell')\in\udV:\ell'=\ell} p_\ude \bbH_k^\ell \bbW_\ude \bbW_\ude^\dag (\bbH_k^\ell)^\dag\right|, 1 \right\} \label{fnf}
\end{align}
for any $\udV\subseteq\udF$.
 The   problem in (\ref{p2ppAlgo})   can now be cast as
\begin{empheq}[box=\widefbox]{align} \label{p2drel}
\max_{\udV\subseteq\udF} &~ f(\udV)~~~\mathrm{s. to} ~~\sum_{(\ude,\ell')\in\udV:\ell'=\ell}r_{\ude} \leq d,  \sum_{(\ude,\ell')\in\udV:\ell'=\ell}p_{\ude}  \leq P, ~~\;\forall\;\ell=1,\ldots,L.
\end{empheq}
%

The following proposition states an important property possessed by the set function $f(\cdot)$.

\begin{proposition}\label{prop:newsf}
The function $f(\cdot)$ in \eqref{fnf} is a monotonic submodular set function over the ground set $\udF$.
\end{proposition}
\proof The proof is given in Appendix   \ref{app:newsf}.

In order to take advantage of recently developed submodular function maximization algorithms, the rank and power constraints  in \eqref{p2drel} need to be cast into the form of linear packing (knapsack) constraints. This can be accomplished readily by associating    each element in $\udF$ with a unique index in $1,\cdots,|\udF|$, where we note $|\udF|=L|\udE|$, and for each subset $\udV \subseteq \udF$ letting $\bbx_{\udV}$ denote a binary ($\{0,1\}$) valued vector of length $L|\udE|$ that has ones in positions indexed by the indices corresponding to elements in $\udV$ and zeros elsewhere. Then, the $2L$ constraints in \eqref{p2drel} can be represented as $\bbA\bbx_{\udV}\leq\bbb$, where $\bbA$ is a $2L\times L|\udE|$ matrix whose rows correspond to the constraints and whose columns correspond to the elements of $\udF$.  Thus, \eqref{p2drel} can be re-cast as
\begin{empheq}[box=\widefbox]{align} \label{p2drel2}
\max_{\udV\subseteq\udF} &~ f(\udV)~~~\mathrm{s. to} ~~\bbA\bbx_{\udV}\leq\bbb.
\end{empheq}

There are some parameters from $\bbA$ and $\bbb$, that are worth  pointing out and are important for characterizing the approximation factors of the algorithms proposed below. First, $\delta := \min_{m,j} \{b_m/A_{m,j}: A_{m,j}>0\}$ is defined as the \emph{width} of all the packing constraints and note that $\delta\geq 1$. Secondly,    there are only $k=2$ non-zero entries per column of $\bbA$. Thus,  the constraints in \eqref{p2drel2} are column-sparse ones and hence \eqref{p2drel2} can be solved using  an algorithm for submodular maximization under column sparse knapsack constraints,   proposed in \cite{bansal_icipco10}. This algorithm (whose complexity scales polynomially in $|\udE|L$) involves randomized rounding combined with alteration  and guarantees an constant approximation factor  which does not depend on   $L$.
%
However, since that randomized   algorithm is computationally demanding to implement, here we employ an algorithm from \cite{gamzu_manu11}, designed for approximately solving submodular maximization under arbitrary knapsack constraints,  instead   and tabulate  it in Algorithm \ref{alg:p2greedy}. Note that in Algorithm \ref{alg:p2greedy} we assume that $\ccalM(\udv)$ returns the index corresponding to any element $\udv\in\udF$. Further,
an expansion step (which is important to establish a performance guarantee for Algorithm \ref{alg:p2DCB}) is added  as the last step of Algorithm \ref{alg:p2greedy}. To explain this expansion, we first define
 $\tilde{\ccalC}_d$ to be a subset of $\ccalC_d$ comprising of all maximal precoding matrix codewords in $\ccalC_d$, i.e., no precoding matrix in $\tilde{\ccalC}_d$ can be expanded by adding any element from $\udE$ without violating the rank or power constraints. Then, in the last step in Algorithm \ref{alg:p2greedy} we ensure that $\hat{\udV}$  is expanded so that each one of its corresponding set of $L$ codewords $\{\bbW^{(\ell)}\}_{\ell=1}^L$ lies in $\tilde{\ccalC}_d$. Notice that
 since each $R_k^\ell(.)$ is a monotonic set function over $\udE$, any arbitrary expansion will improve the value of the objective function.


\begin{algorithm}[htb]
\caption{: To approximately solve \eqref{p2drel2}. Input the channel matrices $\{\bbH_k^\ell\}$, $\bbA$   and $\bbb$ as in \eqref{p2drel2} and an update factor $\lambda \in \mathbb R_{+}$. Output a subset $\hat{\udV}$.} 
\begin{algorithmic} \label{alg:p2greedy}
\STATE Initialize $\udV' = \emptyset$;
\FOR{$m=1,\ldots,2L$}
    \STATE Set the variable $\omega_{m} \leftarrow 1/b_{m}$.
\ENDFOR
\WHILE{$\sum_{m=1}^{2L} b_m \omega_m \leq \lambda$, and $\udV'\neq \udF$}
    \STATE Find  $\hat{\udv}=\arg\min_{\udv\in\udF \setminus \udV'}\left[\sum_{m=1}^{2L} A_{m,\ccalM(\udv)}\omega_m/(f(\udV'\cup{\ccalM(\udv)})-f(\udV'))\right]$ and let $\hat{i}=\ccalM(\hat{\udv})$ denote its corresponding index.
    \STATE Update $\udV' \leftarrow \udV' \cup{\hat{\udv} }$.
    \FOR{$m=1,\ldots,2L$}
    \STATE Update $\omega_m \leftarrow \omega_m \lambda ^{A_{m,\hat{i}}/b_m}$.
\ENDFOR
\ENDWHILE
\IF{$\bbA\bbx_{\udV'} \leq \bbb$}
    \STATE Set $\hat{\udV}=\udV'$.
\ELSIF{$f(\udV'\setminus \ccalM(\hat{i}))\geq f(\ccalM(\hat{i})) $}
    \STATE Set $\hat{\udV}=  \udV'\setminus \ccalM(\hat{i}) $.
\ELSE
    \STATE Set $\hat{\udV}=  \ccalM(\hat{i}) $.
\ENDIF
\end{algorithmic}
Expand $\hat{\udV}$ if needed and output it.
\end{algorithm}

The following result, which holds even when no expansion is employed in the last step of Algorithm \ref{alg:p2greedy},  follows upon invoking Theorem 1 from \cite{gamzu_manu11}.

\begin{lemma}\label{prop:p2greedy}
Algorithm \ref{alg:p2greedy} is a deterministic polynomial-time algorithm that attains an approximation ratio of $~\Omega(1/(2L)^{1/\delta})$. In other words, its final output $\hat{\udV}$ is feasible, i.e.,  $\bbA\bbx_{\hat{\udV}}\leq\bbb$
 and also achieves a constant approximation guarantee
\begin{align}
f(\hat{\udV}) \geq  \Omega\left(1/(2L)^{1/\delta}\right)  ~\max_{\udV:\bbA\bbx_{\udV}\leq\bbb } f(\udV).
\end{align}
\end{lemma}

Before we establish a performance guarantee for Algorithm \ref{alg:p2DCB} we offer the following result which will be invoked later.

\begin{proposition}\label{prop:reformD}
The problem in (P2D)   can be further constrained without loss of optimality by enforcing that each precoder used must lie in the set $\tilde{\ccalC}_d$ and no more than $\lceil\frac{\Theta}{\min_k\Delta_k}\rceil$ scheduling intervals can employ an identical set of maximal precoding matrix codewords.
\end{proposition}

We are now ready to establish the performance guarantee for Algorithm \ref{alg:p2DCB}. The proof of Proposition \ref{prop:reformD}  as well as the one below are given in Appendix \ref{app:reformD}.

\begin{proposition}\label{prop:reformD2}
The solution returned by Algorithm \ref{alg:p2DCB} guarantees a weighted sum delay that is  no greater than
 $\Gamma\ln(1/\epsilon)$ times that of the optimal solution to (P2D), where $\Gamma$ is a fixed constant and the scalar $\epsilon$ is dependent on the  input set of channel matrices, as
 \begin{align}
\epsilon=\min_{k\in\{1,\cdots,K\}}\;\;\min_{\{\bbW^{(\ell)}\in\tilde{\ccalC}_d\}_{\ell=1}^L:\sum_{\ell=1}^LR_k^\ell(\bbW^{(\ell)})>0}\left\{\sum_{\ell=1}^LR_k^\ell(\bbW^{(\ell)})\right\}
\end{align}
\end{proposition}

Note that $\epsilon$ represents the smallest positive rate   that can be achieved by using   maximal codewords over a scheduling interval.


%
%

\vspace{-5mm}


\section{Numerical Examples}\label{sec:sim}

In this section, the effectiveness of the proposed algorithms is shown through numerical tests, where independently and identically distributed Rayleigh fading between the BS and each user is assumed.

\begin{figure}[htb!]
\begin{center}
\centerline{\epsfig{file=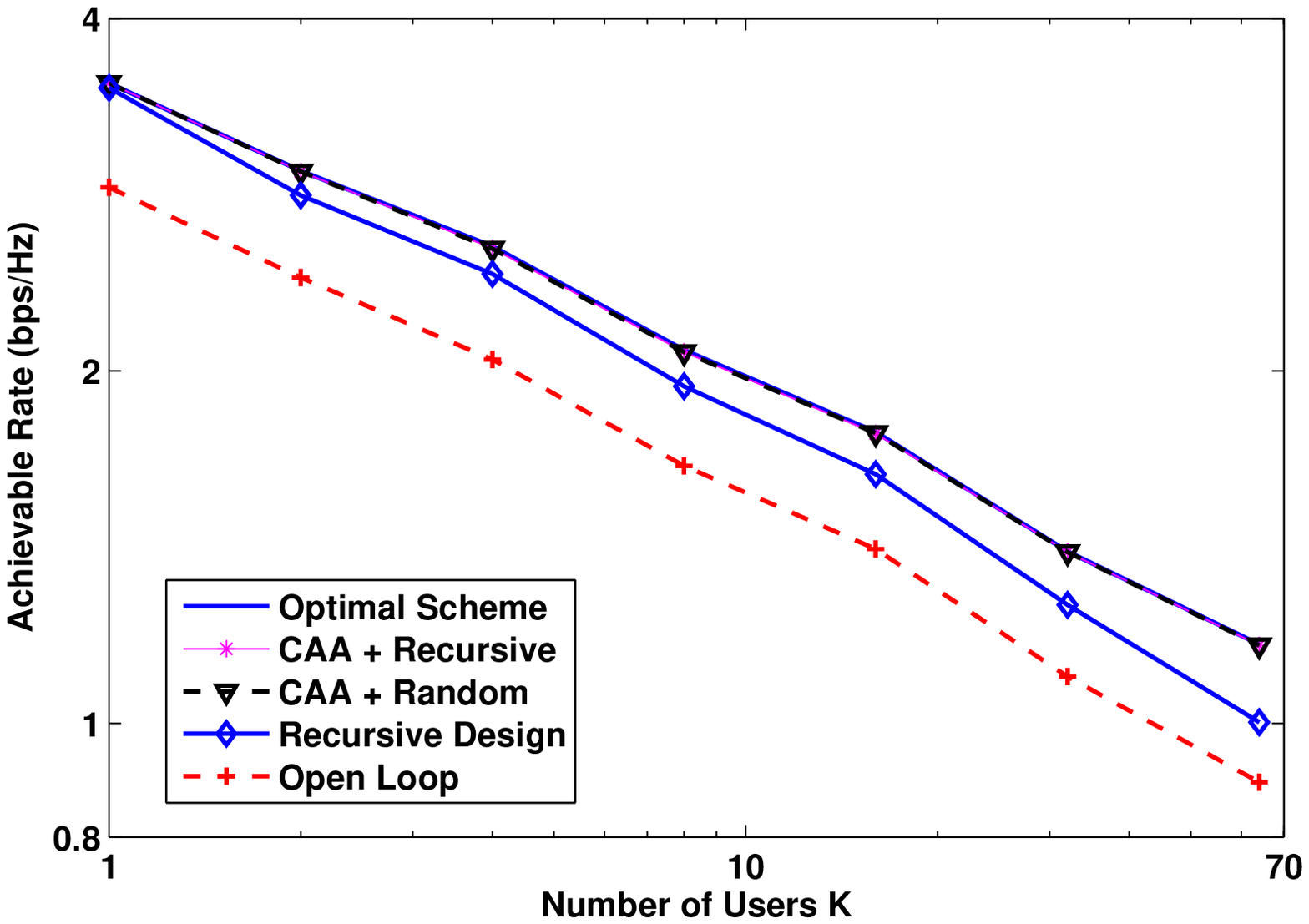,width=.8\linewidth}}
\centerline{\small (a)}
\centerline{\epsfig{file=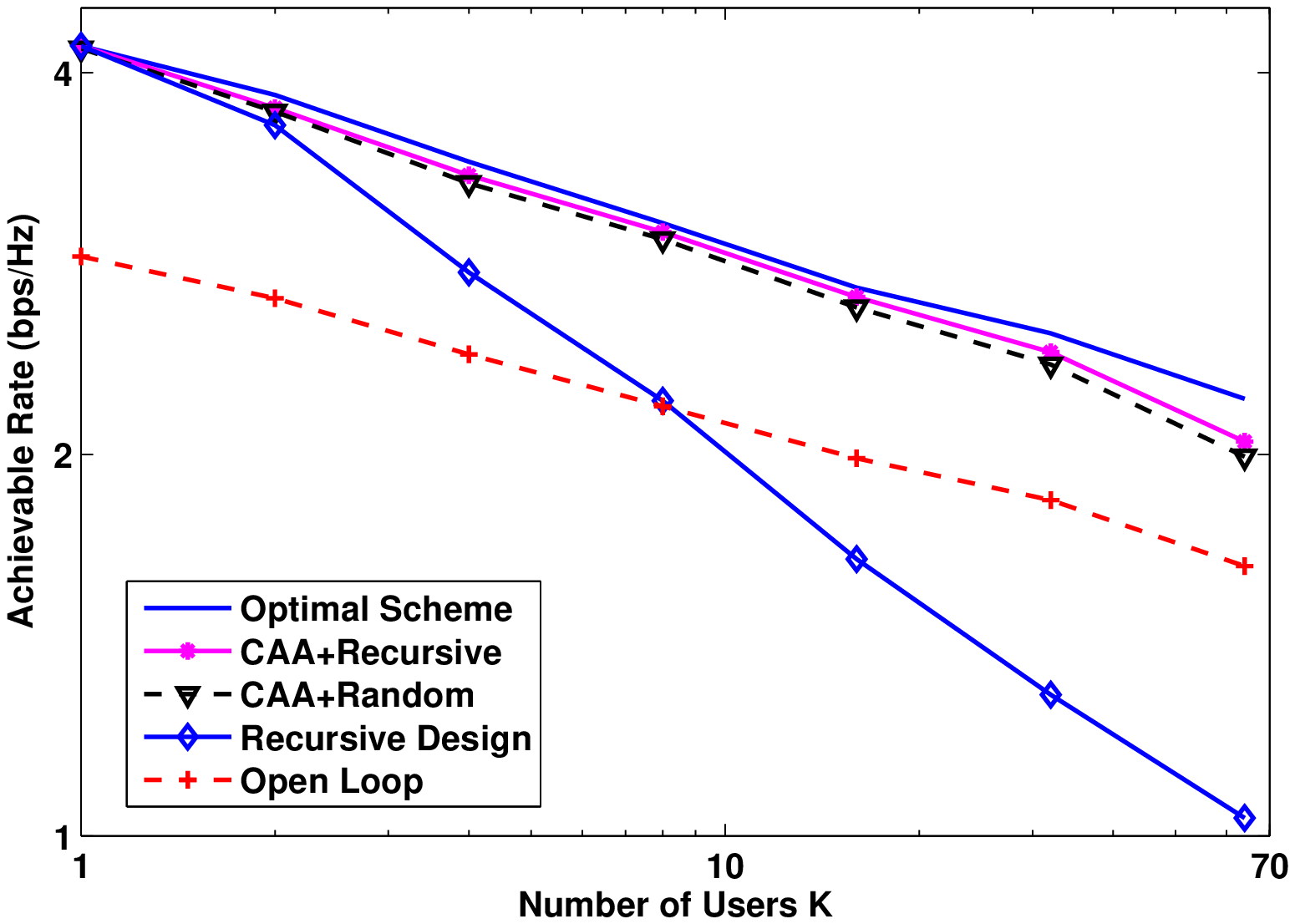,width=.8\linewidth}}
\centerline{\small (b)}
\caption{The maximum achievable rates, with (a) $M=2$ and (b) $M=4$ transmitting antennas and $N=1$ receive antenna, versus number of users $K$ for different schemes  ($P=10$ and $d=2$).}
\label{fig:misoc}
\end{center}
\end{figure}

\noindent \emph{Test Case 1:} The MISO channel from the BS to each single antenna user is considered with the number of transmitting antennas being $M=2$ and  $M=4$, respectively. Fig. \ref{fig:misoc} plots the maximum achievable rate of different schemes with respect to (wrt) the number of users $K$, which increases from 1 to 64, both in the logarithmic scales. The power budget is set to $P=10$, such that the equivalent transmit signal-to-noise ratio (SNR) is 10dB. The proposed CAA algorithm with number of streams $d=2$ is compared with three other schemes. The optimal scheme with number of streams $d=M$ is obtained by solving a semi-definite program (SDP) using SeDuMi \cite{sedumi}, whereas the open-loop precoder refers to the case where $\bbW$ is a scaled identity matrix. Moreover, the recursive design proposed in \cite{kim_rws08} by setting $d=2$ is also compared, and used to initialize the CAA algorithm besides a random initialization.
Note that both the recursive design and CAA are constrained by $d=2$ so that neither can achieve the optimal scaling when $M=4$. Nevertheless, even in this case the CAA algorithm with both initializations keeps on exhibiting  near-optimal performance, especially considering the fact that the optimal scheme with $d=M=4$ provides the non-achievable upper bound.  This clearly shows the near-optimal performance of the proposed CAA algorithm and its insensitivity to initializations.

\vspace{2mm}

\begin{figure}[htb!]
\begin{center}
\centerline{\epsfig{file=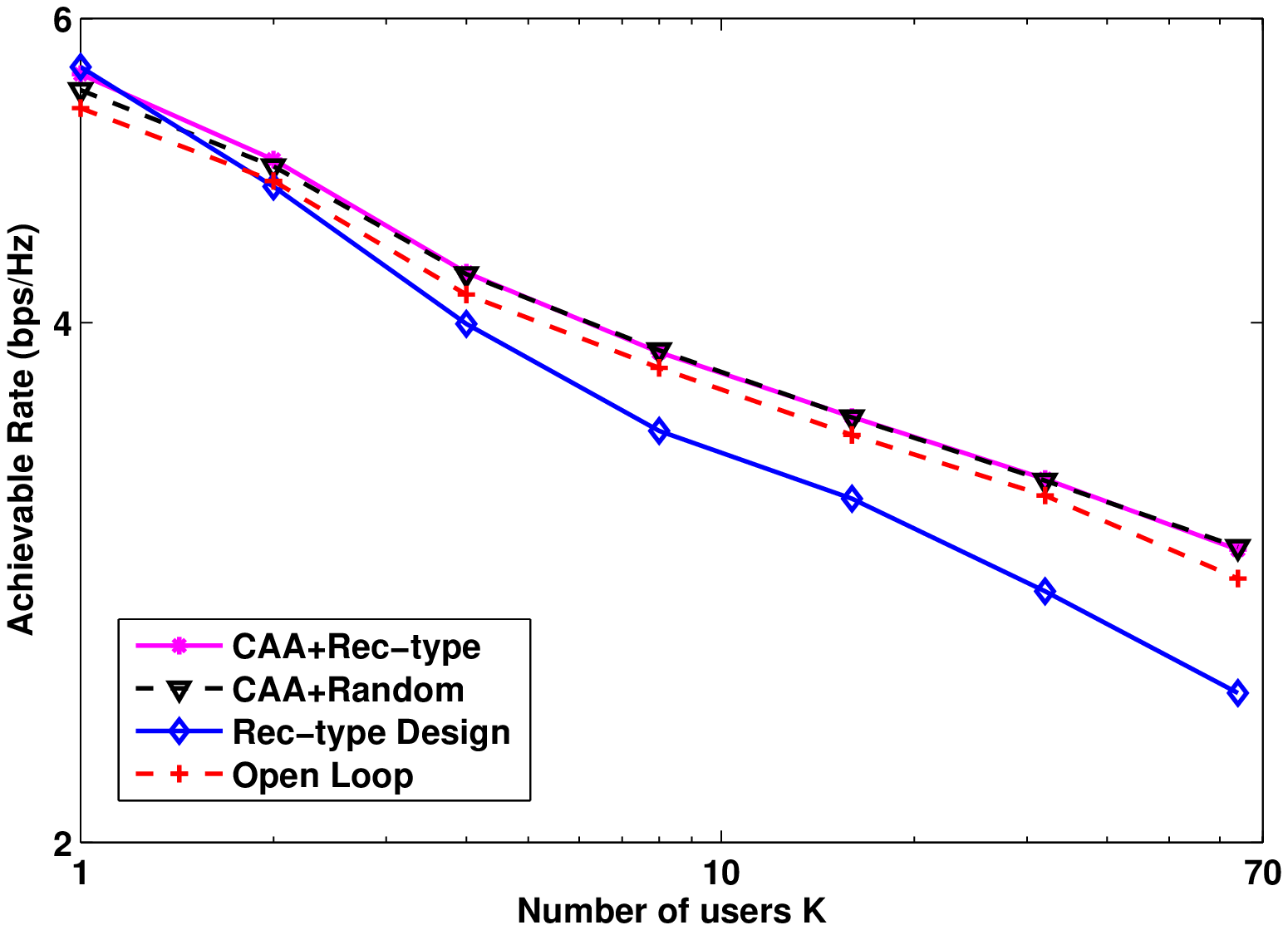,width=.8\linewidth}}
\centerline{\small (a)}
\centerline{\epsfig{file=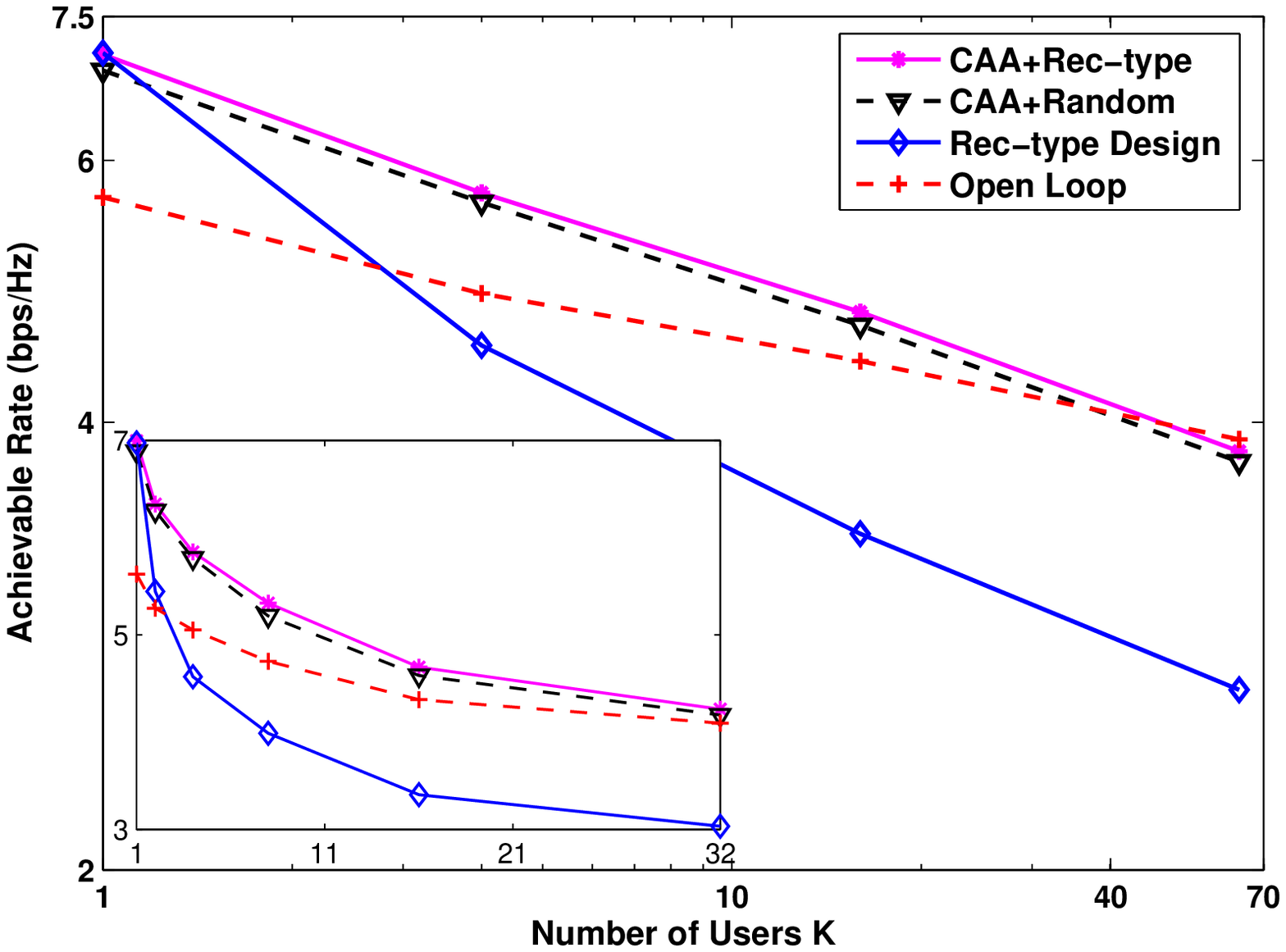,width=.8\linewidth}}
\centerline{\small (b)}
\caption{The maximum achievable rates, with (a) $M=2$ and (b) $M=4$ transmitting antennas and $N=2$ receive antennas, versus number of users $K$ for different schemes  ($P=10$ and $d=2$).}
\label{fig:mimoc}
\end{center}
\end{figure}

\noindent \emph{Test Case 2:} The system settings are the same as those in Test Case 1, except for the number of receive antennas $N_k=2,\forall\;k$ so that each user has a MIMO channel. For this case, the optimal precoder design is no longer an SDP problem, but the open-loop scheme still has the same scaling wrt $K$ as the optimal one. As seen in Fig. \ref{fig:mimoc}, the CAA algorithm fails to achieve the optimal scaling when $M=4$.  However, inspite of being constrained by $d=2$, CAA algorithm still outperforms the open-loop one with $d=4$ when the number of users is less than 32, and its advantage over an intuitive extension of recursive design (referred to as Rec-type design) wherein the channel matrix to the worst user is used as the transmit precoder after appropriate scaling, becomes more evident. For clarity a sub-figure in the linear scale has  also been plotted.


\vspace{2mm}

\begin{figure}[htb!]
\begin{center}
\centerline{\epsfig{file=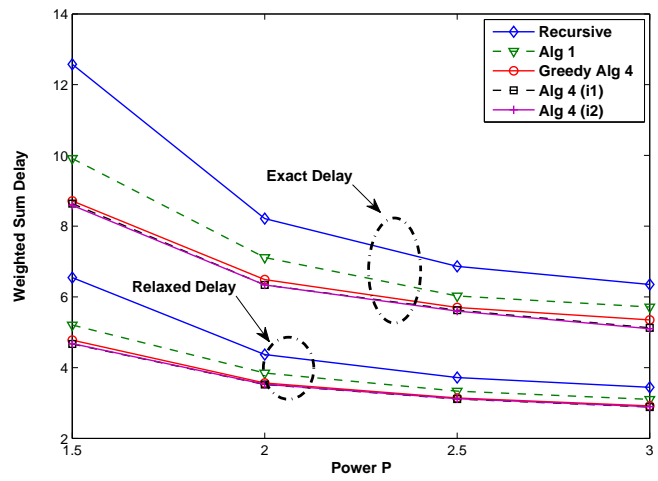,width=.6\linewidth}}
\centerline{\small (a)}
\centerline{\epsfig{file=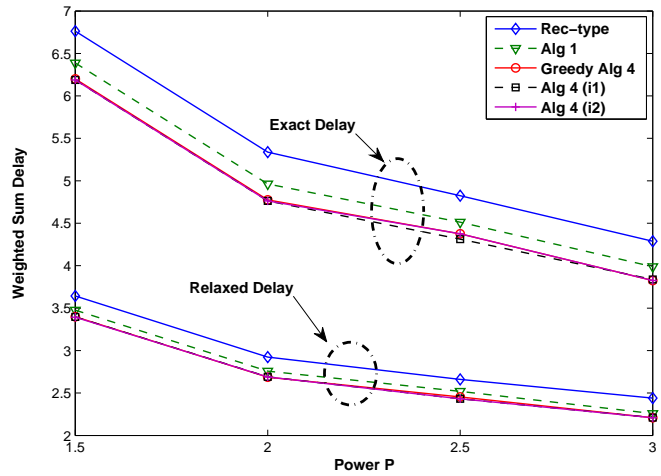,width=.6\linewidth}}
\centerline{\small (b)}
\centerline{\epsfig{file=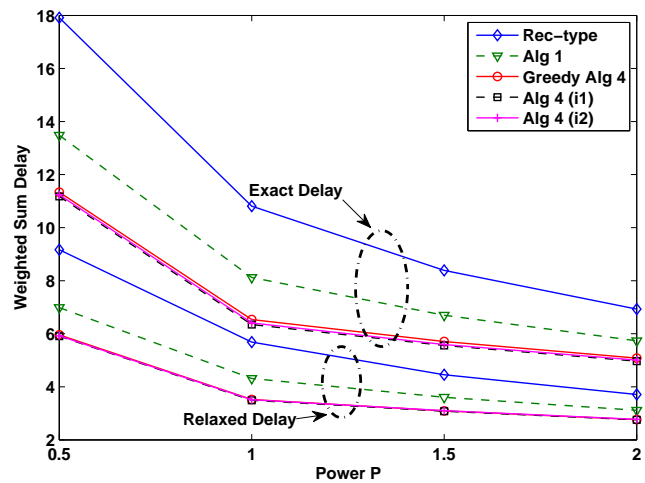,width=.6\linewidth}}
\centerline{\small (c)}
\caption{The weighted sum delay with $M=4$ transmit antennas and (a) $N=1$ receive antenna with equal user weights and (b) $N=2$ receive antennas with equal user weights and (c) $N=2$ receive antennas with unequal user weights, versus the transmit power for different schemes.}
\label{fig:p2c}
\end{center}
\end{figure}

\noindent \emph{Test Case 3:}
The minimum weighted sum delay problem (P2) is now considered over a system in which the BS has $M=4$ antennas and there are $K=8$ users and the transmit precoders are constrained to have rank no greater than two (i.e., $d=2$). In addition, the rate threshold $\Theta$ is set to be 10, while for simplicity the number of orthogonal slots per interval is set as $L=1$. The CAA-based Algorithm \ref{alg:p2ad} which jointly optimizes all the precoding matrices is compared with two other schemes, where in each interval (or equivalently here in each slot) the same precoder is employed and this precoder in turn is either obtained by the recursive design or by Algorithm \ref{alg:p1ad}, respectively, as detailed in Test Cases 1 and 2. The greedy Algorithm \ref{alg:p2ad} corresponds to the reduced complexity scheme which involves solving for $\bbW^\tdt$ after fixing all precoders prior to slot $\tdt$. Further, two approaches for initializing Algorithm \ref{alg:p2ad} are considered: (i1) upon incrementing to slot $\tdt$ augmenting with $\bbW^\tdt = \hhatbbW$ while fixing all precoders obtained prior to this slot; and (i2) at each slot increment simply setting $\bbW^\tau = \hhatbbW$, $\forall 1\leq \tau\leq \tdt$, where $\hhatbbW$  is the solution obtained using Algorithm \ref{alg:p1ad}. Both the per-user MISO channel ($N_k=1,\;\forall\;k$) and the MIMO channel ($N_k=2,\;\forall\;k$) are considered with uniform weights $\mu_k=1/K,\forall\; k$. In addition, unequal user weights are also considered for the $N_k=2$ case by setting $\mu_\kappa =0.9$ for the user $\kappa:= \arg \min_{k} R_k$ as determined by the solution of Algorithm \ref{alg:p1ad}, and $\mu_k = 0.1/(K-1)$ for any other user $k\neq \kappa$.  The exact weighted sum delay which is the objective in (P2) and the relaxed one associated with (P2$'$)  are plotted versus the  power budget (per slot) $P$, in Fig. \ref{fig:p2c}. Clearly, the joint optimization schemes of Algorithm \ref{alg:p2ad} yield improvement over the other  ones, particularly so when unequal weights assigned, as expected. Meanwhile, the curves of greedy Algorithm \ref{alg:p2ad} are quite close to the ones of the original Algorithm \ref{alg:p2ad}, which greatly advocates the use of the reduced complexity scheme in practice.
Interestingly, the relaxed delay curves exhibit the same relative behavior as their exact delay counterparts, which justifies using the relaxed problem (P2$'$) to design transmit precoders that reduce the weighted sum delay.


\vspace{2mm}

\begin{figure}[!t]
\centering
\includegraphics[width=3.7in]{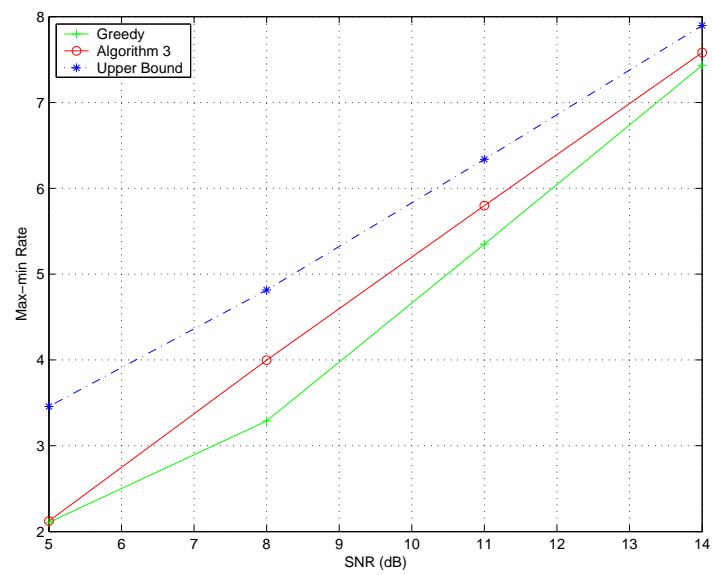}
\caption{The maximum achievable rates with  $M=4$  transmit  antennas and $N=2$ receive antennas, versus transmit SNR for different schemes.}
\label{fig:mimod}
\end{figure}

\noindent \emph{Test Case 4:} We now examine optimization using the discrete codebook $\ccalC_d$.   We consider the rate   optimization in (\ref{p1dss}) over a  system with five users,  with  $N_k=2,\;\forall\;k$ receive antennas  and where the base station has $M=4$ transmit antennas.  The rank one LTE codebook comprising of $16$ unit-norm vectors \cite{3gpp_rel8} formed the base codebook $\ccalW$ and for each codeword an identical set of four power levels is allowed, which together specify the set ground set $\udE$. In Fig. \ref{fig:mimod}
 we plot the achieved throughputs   for different values of transmit SNR. In particular, we have plotted the throughput upper bound obtained obtained by solving (\ref{eq:UBmaxmin}),  as well as that yielded by Algorithm \ref{alg:bis} when the latter is invoked with $\delta=0,\epsilon=.08$ along with its practical refinements discussed in Section \ref{sec:dist_maxinst}. For comparison, we also plot the throughput   yielded by a simple greedy algorithm, which  at each step selects the element from $\udE$ yielding the largest increase in the instantaneous rate subject to the transmit power constraint. Note that at moderate values of SNR  Algorithm \ref{alg:bis} yields a good improvement over the simple greedy baseline. The gains are lower at  high SNRs since in that regime the transmit power constraint becomes increasingly irrelevant (i.e., most of the codebook can be selected). We emphasize that the upper bound which relaxes the binary-value constraints need not be achievable, particularly at low SNRs.

\section{Conclusions and Future Work}
 \label{sec:conclusions}
We considered the design of linear precoders for multicast by using \emph{instantaneous rate} and \emph{weighted sum delay} as the design criteria. The linear precoders were allowed to be any complex valued matrices subject to rank and power constraints (a.k.a. the continuous codebook case). Alternatively, the linear precoders could be constructed by selecting and concatenating codewords from a given finite codebook of  precoding matrices (a.k.a. the discrete codebook case). For the former case, cyclic alternating ascent (CAA) based algorithms were proposed, whereas for the latter case greedy algorithms that exploit submodularity of the rate function were proposed.
The proposed algorithms were shown to possess certain desirable properties such as satisfying KKT conditions and offering worst-case guarantees. 

The CAA based algorithms offer good performance but their complexities can be deemed high for some implementations,  since they involve solving an SOCP in each step. An interesting avenue for future work would be to determine whether explicit solutions   can be obtained for special instances and then leverage them. On the other hand, the greedy algorithms for the discrete codebook case are simple to implement. However, the performance guarantee obtained for the weighted sum delay minimization might be weak and the design of approximation algorithms with better guarantees is an open problem.


Furthermore, recall that the quasi-static assumption adopted for the weighted sum delay minimization problem  allowed us to use any arbitrary number of scheduling intervals to ensure that the threshold for each user is achieved. In problems where a strict limit on the number of intervals is present, we would require an admission control module to select a multi-cast group of users and/or to set an appropriate threshold to ensure that decoding at all users can be achieved. Extending our proposed techniques   to design such a module is an interesting open problem. Finally, developing   robust versions of the results developed in this paper, by adopting a bounded CSI error model (as in \cite{vucic2009,jose_nectr10}) is also an interesting problem. While such an extension is not difficult for the continuous codebook case, its discrete counterpart seems challenging since the submodularity property may no longer hold for the worst-case (over all error realizations) per-user rate.

\bigskip

\appendices

\section{Proof of Proposition \ref{prop:p1adconvg}}\label{app:cca}
To proceed, define the objective for the inner minimization in \eqref{p1c} as
\begin{align}
r_k\left(\bbW, \bbG_k,\bbS_k\right)= -\tr[\bbS_k\bbE_k(\bbG_k,\bbW)] + \log | \bbS_k |+d,\;\;1\leq k\leq K
\end{align}
and the one for the outer maximization as
\begin{align}
g\left(\bbW, \{\bbG_k,\bbS_k\}\right)= \min_{k=1,\ldots,K} ~r_k\left(\bbW, \bbG_k,\bbS_k\right).
\end{align}
Moreover, let $i \in \mathbb Z_{+}$ be the iteration index for the while-loop in Algorithm \ref{alg:p1ad}, and it is initialized with the input precoder $\bbW(0)$. Further, denote the maximal objective values achieved before and after the precode update at the $i$-th iteration as
\begin{align}
g_i &= g\left(\bbW(i-1), \left\{\bbG_k(i), \bbS_k(i)\right\}\right), \nonumber \\
\zeta_i &= g\left(\bbW(i), \left\{\bbG_k(i), \bbS_k(i)\right\}\right) ~~\forall i \in \mathbb Z_{+}~. \label{zetai}
\end{align}
The ascent nature of the iterations in Algorithm \ref{alg:p1ad} ensures the sequence $\{g_i\}$ is monotonically non-decreasing and hence convergent, and also $g_i \leq \zeta_i \leq g_{i+1}$, $\forall i\in  \mathbb Z_{+}$ . Due to the boundedness of $\{\|\bbW(i)\|\}$ ensured by the norm constraint in \eqref{p1c}, there exists a   subsequence $\ccalI$ such that $\bbW(i) \rightarrow \bbarbbW$, $i\in\ccalI$. Line 3 of Algorithm \ref{alg:p1ad} indicates that $\bbG_k(i+1)$ is obtained from an analytical function of $\bbW(i)$, thus   it follows that for any $k$, $\bbG_k(i+1) \rightarrow \bbarbbG_k$, $i\in\ccalI$. Similar  argument holds for each $\bbS_k(i+1) \rightarrow \bbarbbS_k$, $i\in\ccalI$. Consequently, the convergence for the objective value sequence follows, as
\begin{align}
g_{i+1} \rightarrow \bbarg := g\left(\bbarbbW, \{\bbarbbG_k,\bbarbbS_k\}\right),~i\in\ccalI  .
\end{align}
Note that since the sequence $\{g_i\}_{i\in\ccalI}$ converges and it is a subsequence of the convergent sequence $\{g_i\}_{i\in\mathbb Z_{+}}$, we must have that
 $g_i\rightarrow \bbarg,\;i\in\mathbb Z_{+}$. Further, the monotonicity of $\{g_i\}_{i\in\mathbb Z_{+}}$ and the relation $g_i \leq \zeta_i \leq g_{i+1}$ ensures that  $\zeta_i\rightarrow \bbarg,\;i\in\ccalI $ as well as $\zeta_{i+1}\rightarrow \bbarg,\;i\in\ccalI $.

Next, we want to show that $\left(\bbarbbW, \{\bbarbbG_k,\bbarbbS_k\}\right)$ constitutes a fixed point for the CAA iterations. Since the updates of $\bbG_k$ and $\bbS_k$ are both closed-form for any $k$, it is easy to see that
\begin{align}
\bbarbbG_k &= \left(\bbH_k\bbarbbW\bbarbbW^\dag \bbH_k^\dag + \bbI_{N_k}\right)^{-1} \bbH_k \bbarbbW \label{barG}\\
\bbarbbS_k &=  \bbarbbW^\dag \bbH_k^\dag\bbH_k\bbarbbW + \bbI_{d}\label{barS}.
\end{align}
Thus, it remains to prove that $\bbarbbW \in \ccalW(\{\bbarbbG_k,\bbarbbS_k\})$, where the later represents the optimal solution set of the SOCP problem \eqref{p1cW} given the inputs $\bbG_k = \bbarbbG_k$ and $\bbS_k = \bbarbbS_k$. Recall that the subsequence $\{\zeta_{i+1}\}_{i\in\ccalI}$ converges to the limit point $\bbarg$. 
Therefore, if it can be shown that
\begin{align}
\zeta_{i+1} 
\rightarrow g\left( \bbU, \{\bbarbbG_k,\bbarbbS_k\}\right), ~i\in\ccalI, \label{zetaconvg}
\end{align}
for some $\bbU\in \ccalW(\{\bbarbbG_k,\bbarbbS_k\})$, then we can deduce that $g\left( \bbU, \{\bbarbbG_k,\bbarbbS_k\}\right) = \bbarg$ from which it follows that   $\bbarbbW \in \ccalW(\{\bbarbbG_k,\bbarbbS_k\})$. To show \eqref{zetaconvg}, consider the following sequence of functions in $\bbW$
\begin{align}
h_{i+1}(\bbW) :=  g\left(\bbW, \{\bbG_k(i+1),\bbS_k(i+1)\}\right),~~\forall i \in \ccalI
\end{align}
where $\zeta_{i+1} = h_{i+1} (\bbW(i+1)) = \max_{\bbW:\|\bbW\|_F^2\leq P}~h_{i+1}(\bbW)$. Since each function $r_k$ is quadratic in $\bbW$ (with $\bbG_k$ and $\bbS_k$ given), and $h_{i+1}$ is the minimum of a finite number of such $r_k$'s, it can be shown that the sequence of functions $\{h_{i+1}(\cdot)\}_{i\in\ccalI}$ converges point-wise to the following function
\begin{align}
\bbarh(\bbW) :=  g\left(\bbW, \{\bbarbbG_k,\bbarbbS_k\}\right).
\end{align}
Further, since only the compact set defined by the norm ball $\|\bbW\|_F^2\leq P$ is of interest, point-wise convergence in $\{h_{i+1}(\cdot)\}_{i\in\ccalI}$ leads to the uniform convergence; that is, for any $\epsilon >0$, there exists an iteration index $i'\in\ccalI$ such that $|h_{i+1}(\bbW) - \bbarh(\bbW)| \leq \epsilon$, $\forall i \in \ccalI$, $i\geq i'$, and $\|\bbW\|_F^2\leq P$. From this uniform convergence, it holds that
\begin{align}
\nonumber\zeta_{i+1} = \max_{\bbW:\|\bbW\|_F^2\leq P}~h_{i+1}(\bbW) \leq \max_{\bbW:\|\bbW\|_F^2\leq P}\left[\bbarh(\bbW) + \epsilon\right]\\ = \max_{\bbW:\|\bbW\|_F^2\leq P}\bbarh(\bbW) + \epsilon,~~\forall i\in\ccalI, i\geq i',
\end{align}
and similarly
\begin{align}
\nonumber\zeta_{i+1} = \max_{\bbW:\|\bbW\|_F^2\leq P}~h_{i+1}(\bbW) \geq \max_{\bbW:\|\bbW\|_F^2\leq P}\left[\bbarh(\bbW) -\epsilon\right] \\= \max_{\bbW:\|\bbW\|_F^2\leq P}\bbarh(\bbW) - \epsilon,~~\forall i\in\ccalI, i\geq i',
\end{align}
and this leads to the following convergence
\begin{align}
\zeta_{i+1} 
\rightarrow \max_{\bbW:\|\bbW\|_F^2\leq P}\bbarh(\bbW) = g\left( \bbU, \{\bbarbbG_k,\bbarbbS_k\}\right), ~i\in\ccalI, 
\end{align}
for some $\bbU\in \ccalW(\{\bbarbbG_k,\bbarbbS_k\})$,
which is sufficient for claiming \eqref{zetaconvg} and completing the proof that $\left(\bbarbbW, \{\bbarbbG_k,\bbarbbS_k\}\right)$ is a fixed point for the CAA iterations. With $\left(\bbarbbW, \{\bbarbbG_k,\bbarbbS_k\}\right)$ in hand the remaining part of the proposition follows by first noting that
\begin{align}\label{eq:eq1}
\left[-\tr(\bbarbbS_k\bbE_k(\bbarbbG_k,\bbW))+\log | \bbarbbS_k |+d\right]\;\big|_{\bbW=\bbarbbW}= \log|\bbI+\bbH_k\bbW\bbW^{\dag}\bbH_k^{\dag}|\;\big|_{\bbW=\bbarbbW}.
\end{align}
Then,  specializing the gradient formulas in \cite{christen} to our case we get that
\begin{align}
\nabla_{\bbW}\left[-\tr(\bbarbbS_k\bbE_k(\bbarbbG_k,\bbW))+\log | \bbarbbS_k |+d\right]=\bbH_k^{\dag}\bbH_k\bbW\bbE_k(\bbarbbG_k,\bbW)\bbarbbS_k\bbE_k(\bbarbbG_k,\bbW)
\end{align}
and
\begin{align}
\nabla_{\bbW}\log|\bbI+\bbH_k\bbW\bbW^{\dag}\bbH_k^{\dag}|=\bbH_k^{\dag}\bbH_k\bbW (\bbI+ \bbW^{\dag}\bbH_k^{\dag}\bbH_k\bbW)^{-1}
\end{align}
so that
\begin{align}\label{eq:eq2}
\nabla_{\bbW}\left[-\tr(\bbarbbS_k\bbE_k(\bbarbbG_k,\bbW))+\log | \bbarbbS_k |+d\right]\;\big|_{\bbW=\bbarbbW}= \nabla_{\bbW}\log|\bbI+\bbH_k\bbW\bbW^{\dag}\bbH_k^{\dag}|\;\big|_{\bbW=\bbarbbW}
\end{align}
Using (\ref{eq:eq1}) and (\ref{eq:eq2}) we can conclude that $\left(\bbarbbW, \{\bbarbbG_k,\bbarbbS_k\}\right)$ satisfy the KKT conditions of (P1) as well. \myQED

\section{Proof of Proposition \ref{prop:submod1}} \label{app:subm}
 Consider any subsets $\udU\subseteq \udU' \subseteq \udE$ such that $\udU'= \udU \cup \udV$. Note that it suffices to consider $\ude' \in\udE\setminus\udU'$ since the proposition is trivially true for $\ude\in\udU'$. Define a function $f_k(\udA)=\sum_{\ude\in\udA} p_\ude \bbH_k \bbW_\ude \bbW_\ude^\dag \bbH_k^\dag,\;\forall \udA\subseteq \udE$. Then, for any element $\ude'\in\udE\setminus\udU'$   we have
\begin{align}
\nonumber &R_k(\udU'\cup\{\ude'\})  - R_k(\udU')
=\log|\bbI + f_k(\{\ude'\})+f_k(\udU') | -\log|\bbI +  f_k(\udU') |\\
\nonumber =&R_k(\udU\cup\{\ude'\})-R_k(\udU)
 +\log \left|\bbI +\sum_{\ude\in\udV}p_\ude\bbW_\ude^\dag \bbH_k^\dag \left(\bbI + f_k(\{\ude'\})+f_k(\udU) \right)^{-1}\bbH_k \bbW_\ude \right|\\
  &\hspace{22mm} - \log \left|\bbI +\sum_{\ude\in\udV}p_\ude\bbW_\ude^\dag \bbH_k^\dag \left(\bbI+f_k(\udU)\right)^{-1}\bbH_k \bbW_\ude \right|.\label{eq:chainP}
\end{align}
Note that $\left(\bbI + f_k(\{\ude'\})+f_k(\udU) \right)^{-1}\preceq \left(\bbI +  f_k(\udU) \right)^{-1}$, where $\preceq$ denotes the positive semi-definite ordering, since  $f_k(\{\ude'\}),f_k(\udU)$ are both positive semi-definite matrices, from which we can deduce that
\begin{align}
\log \left|\bbI +\sum_{\ude\in\udV}p_\ude\bbW_\ude^\dag \bbH_k^\dag \left(\bbI + f_k(\{\ude'\})+f_k(\udU) \right)^{-1}\bbH_k \bbW_\ude \right|\leq \log \left|\bbI +\sum_{\ude\in\udV}p_\ude\bbW_\ude^\dag \bbH_k^\dag \left(\bbI+f_k(\udU)\right)^{-1}\bbH_k \bbW_\ude \right|.\label{eq:psdO}
\end{align}

Substituting (\ref{eq:psdO}) in (\ref{eq:chainP}) leads to (\ref{Rktsub}). The remaining parts can be readily verified to be true.  \myQED

\section{Proof of Proposition \ref{prop:p1dnp}}\label{app:np}

To show the hardness of the discrete precoder design problem \eqref{p1dss}, consider an instance of the \emph{hitting set problem}, which is among Karp's 21 NP-complete problems \cite{Garey_np79}. Specifically, with a collection of $\tilde{K}$ subsets $\{\ccalS_k\}_{k=1}^{\tilde{K}}$ of a ground set $\udS$, and a positive integer $P'$, the goal is to find whether there exists a \emph{hitting set} $\udS'$ of size $P'$ or less, that is, a subset $\udS'\subseteq \udS$ such that  $|\udS'| \leq P'$ and $\udS' \cap \ccalS_k \neq \emptyset$, $\forall k = 1,\ldots, \tilde{K}$. For convenience, given any element $s \in \udS$, let the indices of those subsets that $s$ belongs to form the set $K(s) \subseteq \{1,\ldots,K\}$, such that $s \in \ccalS_k$, $\forall k \in K(s)$, and $s \notin \ccalS_k$, $\forall k \notin K(s)$. Furthermore, we restrict our attention to instances constrained to satisfy $\tilde{K}=O(|\udS|^{\Delta})$  for any arbitrarily fixed positive integer $\Delta\geq 2$. We note that the hitting set problem remains NP hard even under such restriction \cite{Garey_np79}.

To map this hitting set instance to one instance of \eqref{p1dss}, wlog let $K=\tilde{K}$ and assume that each user $k$ is equipped with one antenna; i.e., $N_k=1$, $\forall k$. Let the number of transmit antennas $M = K$, and set the channel gain vectors $\{\bbH_k\}$ to be orthonormal such that $\bbH_k \bbH_l^\dag = \delta_{k,l}$, where the later represents the Kronecker delta operator. For the discrete codebook $\ccalC_d$, consider a flat power profile for each codeword as $p_{\ude} = 1$, $\forall \ude \in \udE$. Moreover, for any element $s \in \udS$ and its companion set $K(s)$, there exists an element $g(s)=\ude \in \udE$, such that the corresponding codeword has rank 1, in the form of
\begin{align}
\bbW_\ude = \frac{1}{\sqrt{|K(s)|}} \sum_{k\in K(s)} \bbH_k^\dag.
\end{align}
Notice that any element $s$ with $K(s) = \emptyset$ can be included as a special case for this codeword setting, which simply renders the corresponding $\bbW_\ude = \bb0$. Under this codeword definition, the achievable rate at user $k$ as a set function in \eqref{Rktset} becomes
\begin{align}
R_k(\udU) = \log \left|1 + \sum_{\ude\in\udU}  \bbH_k \left(\frac{1}{\sqrt{|K(g^{-1}(\ude))|}} \sum_{l\in K(g^{-1}(\ude))} \bbH_l^\dag\right)
\left(\frac{1}{\sqrt{|K(g^{-1}(\ude))|}} \sum_{l\in K(g^{-1}(\ude))} \bbH_l\right) \bbH_k^\dag\right|, ~~\forall k.
\end{align}
If there exists some $\ude' \in \udU$ such that its corresponding $s'=g^{-1}(\ude') \in\ccalS_k$, then it holds
\begin{align}
R_k(\udU) &\geq \log \left|1 + \bbH_k \left(\frac{1}{\sqrt{|K(s')|}} \sum_{l\in K(s')} \bbH_l^\dag\right)
\left(\frac{1}{\sqrt{|K(s')|}} \sum_{l\in K(s')} \bbH_l\right) \bbH_k^\dag\right|  \nonumber\\
& = \log \left(1+ 1/|K(s')|\right) \geq \log  (1+1/K)
\end{align}
where the last inequality comes from $|K(s')|\leq K$. Otherwise, if for any $\ude' \in \udU$, the corresponding $s' \notin \ccalS_k$, then it can also be shown that $R_k(\udU) = 0$. Therefore, for each set $\ccalS_k$, the set function $R_k(\udU') \geq \log  (1+1/K)$ if the subset $\udS'=\{g^{-1}(\ude):\ude\in\udU'\}\subseteq \udS$ corresponding to all codewords in $\udU'$ intersects $\ccalS_k$, and 0 otherwise. If we assume an optimal solution to the hitting set instance is $\udS^*$ of size no greater than $P'$, then for the corresponding set $\udU^*$ we have $\min_k R_k(\udU^*) \geq \log  (1+1/K)$. For any other $\udS'$ that is not a hitting set, we have the corresponding $\min_k R_k(\udU') = 0$.

Now consider the precoder design problem \eqref{p1dss} under the current settings. Due to the flat power profile, $p_\udU \leq P$ is equivalent to a cardinality constraint   $|\udU| \leq \lfloor P \rfloor$. To establish the connection to the hitting set problem, let $\lfloor P\rfloor = P'$. If there were an algorithm for \eqref{p1dss} with approximation guarantee $  \gamma(|\udE|)$, it would select a set $\udU'$ of size $|\udU| \leq P'$ with
\begin{align}
\min_{k} R_k(\udU') \geq \gamma(|\udE|) \left[\min_k R_k(\udU^*) \right] =\gamma(|\udE|) \log  (1+1/K) >0.
\end{align}
This implies $\min_k R_k(\udU') \geq \log  (1+1/K)$, and thus the subset $\udS'\subseteq \udS$ corresponding to $\udU'$ would be a hitting set. Accordingly, this approximation algorithm would be able to decide, whether there exists a hitting set of size $P'$, contradicting the NP-hardness of the hitting set problem \cite{Garey_np79}. \myQED

\section{Proof of Proposition \ref{prop:p2bis} }\label{app:alg4}
Note that Algorithm \ref{alg:bis} clearly converges and let $\hat{c}$ denote the value of $c$ obtained upon convergence. Invoking Lemma \ref{lem:gpcM} we can conclude that
\begin{align}\label{eq:fireq0}
\bbarR_{\hat{c}}(\hat{\udU})\geq \hat{c}(1-\delta)
\end{align}
   with $p_{\hat{\udU}} \leq  P(1+\ln(1/\delta))$. Further since that value $\hat{c}+\epsilon$ cannot be achieved by Algorithm \ref{alg:gpcM} without exceeding the budget $P(1+\ln(1/\delta))$, from Lemma \ref{lem:gpcM}, (\ref{p1dbi2}) and (\ref{p1dbi}) we can also deduce that $p_{\hat{\udU}_{\hat{c}+\epsilon}}>P$ so that
\begin{align}\label{eq:fireq}
\max_{\udU:p_\udU \leq P} \min_{k} R_k(\udU)< \hat{c}+\epsilon.
\end{align}
Next, expanding $\bbarR_{\hat{c}}(\hat{\udU})=(1/K) \sum_{k=1}^K \min\{R_{k}(\hat{\udU}),\hat{c}\}$ and using (\ref{eq:fireq0}), we can show via contradiction that we must have
\begin{align}\label{eq:fireq2}
  R_k(\hat{\udU})\geq \hat{c}(1-K\delta),\;\forall\;1\leq k\leq K.
\end{align}
(\ref{eq:fireq}) and (\ref{eq:fireq2}) together prove the proposition. \myQED

\section{Proof of Proposition \ref{prop:propoFK} }\label{app:propoFK}
We first assume that for the given input channel set there exists a set of feasible precoding matrices $\{\breve{\bbW}^{(\ell)}\}_{\ell=1}^L$ such that
$\sum_{\ell=1}^LR_k^\ell(\breve{\bbW}^{(\ell)})/\Theta>\Delta$, for all users $1\leq k\leq K$ and for some $\Delta>0$.
Note that  this assumption is not satisfied only if one or more users have mutually orthogonal input channels, i.e. $\sum_{\ell=1}^L(\bbH_k^{\ell})^\dag\bbH_j^\ell=\bb0$ for some $k\neq j$. In that case users can be partitioned into multiple groups with each group satisfying the aforementioned assumption and the arguments given below can be used separately over each group. Then, since $\Theta$ is finite, a feasible solution to ensure that each user decodes the common message  is to repeat $\{\breve{\bbW}^{(\ell)}\}_{\ell=1}^L$ over $\lceil\frac{\Theta}{\Delta}\rceil$ scheduling intervals which then yields a finite value for the objective function in (P2$'$), denoted henceforth by $G$. Letting $\{\bbW_{\rm opt}^\tau\}$ be any optimal solution, we can deduce that the optimal objective function value for (P2$'$) yielded by it is clearly finite and no greater than $G$. By contradiction, it can then be argued that  for each user $k$, $\sum_{\tau=1}^{Lt}R_k^\tau(\bbW_{\rm opt}^{\tau})/\Theta\geq 1-\Delta$ for all $t\geq\frac{G}{\Delta\min_k\mu_k}$. Then, since  $\{\breve{\bbW}^{(\ell)}\}_{\ell=1}^L$ achieves a normalized rate no less than $\Delta$ for each user in a scheduling interval, invoking the optimality of $\{\bbW_{\rm opt}^\tau\}$ we must have that $\sum_{\tau=1}^{Lt}R_k^\tau(\bbW_{\rm opt}^{\tau})/\Theta\geq 1$ for all $t>\hat{t}=1+\frac{G}{\Delta\min_k\mu_k}$. Consequently, without loss of optimality  the given optimal solution can be truncated after $L\hat{t}$.\myQED

\section{Proof of Proposition \ref{prop:p2adconvg} }\label{app:p2adconvg}
Suppose $\tilde{t}$ is the value for the number of scheduling intervals returned upon termination of the while$-$do loop and let
$\{\bbW^\tau\}_{\tau=1}^{L\tilde{t}}$ denote the iterate returned by it. Then, using arguments similar to those made to prove Proposition \ref{prop:p1adconvg} it can be shown that $\{\bbW^\tau\}_{\tau=1}^{L\tilde{t}}$  is a stationary point of (P2$''$) (evaluated for that $\tilde{t}$). Thus, $\{\bbW^\tau\}_{\tau=1}^{L\tilde{t}}$ must be feasible, i.e., $\bbW^\tau\in\ccalC_c,\;\forall\;\tau$ and also satisfy the other KKT conditions for (P2$''$).
Let $\bbA^\tau_k(\bbW^\tau), \bbB(\bbW^\tau)$ denote the derivatives of $R_k^\tau(\bbW^{\tau})$ and $\|\bbW^\tau\|_F^2$ (with respect to the precoding matrix argument)  evaluated at $\bbW^\tau$, respectively. Further, let
$C_k=\max\{t\in\{0,1,\cdots,\tilde{t}\}:\sum_{\tau=1}^{tL}R_k^\tau(\bbW^{\tau})<\Theta\},\;\forall\;k$, where we note that $C_k=0$ if $\sum_{\tau=1}^{L}R_k^\tau(\bbW^{\tau})\geq \Theta$.
Then invoking the KKT conditions for (P2$''$), after some manipulations we can deduce that there must exist non-negative scalars 
 $\delta^\tau,1\leq\tau\leq L\tilde{t}$ such that
 \begin{subequations}\label{p2ppkkt}
\begin{align}
\sum_{k:C_k\geq\lceil\frac{\tau}{L}\rceil}\bbA^\tau_k(\bbW^\tau)\mu_k\left(C_k-\lceil\frac{\tau}{L}\rceil+1\right)=\delta^\tau \bbB(\bbW^\tau),\;\;1\leq\tau\leq L\tilde{t}.
  \label{p2ppkkta}
\end{align}
\end{subequations}
 Clearly using this $\{\bbW^\tau\}_{\tau=1}^{L\tilde{t}}$, the accumulated rate of each user $k$ after $\tilde{t}$ scheduling intervals is no less than $\Theta-\Delta_k$. We only consider the case where  at-least one user's accumulated rate is less than $\Theta$ since the remaining one can be proved in a similar manner. Then, letting $T=\max\{\hat{t},\tilde{t}+1\}$, where we recall $\hat{t}$ was implicitly defined in Proposition \ref{prop:propoFK}, we  consider the KKT conditions for the following problem
 \begin{empheq}[box=\widefbox]{align} \label{p2relA}
  \min_{\{\bbW^\tau\in\ccalC_c\}} ~~ \sum_{k=1}^K ~\sum_{t=1}^{T}\mu_k\left[ 1-  \min \left\{\textstyle \sum_{\tau=1}^{Lt} R_k^\tau(\bbW^\tau)/\Theta, 1\right\}\right]. ~~~~~
\end{empheq}
  Note that any optimal solution of (P2$'$) (truncated without loss of optimality after interval  $T$) must satisfy the KKT conditions for (\ref{p2relA}).
 Now consider the augmented set $\{\hat{\bbW}^\tau\}_{\tau=1}^{LT}$, where
 \begin{eqnarray}\label{p3relA}
  \nonumber \hat{\bbW}^\tau  &=& \left\{\begin{array}{c}
  \bbW^{\tau} ,\;\; \;{\rm If}\; 1\leq\tau\leq L\tilde{t} \\
  \breve{\bbW}^{(\tau-L\tilde{t})},\;\;\;{\rm Else If}\;L\tilde{t}+1\leq\tau\leq L(1+\tilde{t})\\
  {\bf 0}\;\;\;\;\;\;{\rm Otherwise}\end{array}\right.
 \end{eqnarray}
  Letting
  $\hat{C}_k=\max\{t\in\{0,1,\cdots,T\}:\sum_{\tau=1}^{tL}R_k^\tau(\hat{\bbW}^{\tau})<\Theta\},\;\forall\;k$, a key observation is that $\hat{C}_k=C_k\leq \tilde{t},\;\forall\;k$. This fact along with (\ref{p2ppkkta}) allows us to conclude
   that
   \begin{subequations}\label{p2ppkktL}
\begin{align}
\sum_{k:\hat{C}_k\geq\lceil\frac{\tau}{L}\rceil}\bbA^\tau_k(\hat{\bbW}^\tau)\mu_k\left(\hat{C}_k-\lceil\frac{\tau}{L}\rceil+1\right)=\delta^\tau \bbB(\hat{\bbW}^\tau),\;\;1\leq\tau\leq LT,
  \label{p2ppkktLb}
\end{align}
\end{subequations}
which suffices to satisfy the KKT conditions for (\ref{p2relA}) and hence those for (P2$'$).  \myQED

\section{Proof of Proposition \ref{prop:newsf} }\label{app:newsf}



The monotonicity of $f(.)$ can be readily verified. Consider the set function $g_{k}: 2^\udF \rightarrow \mathbb R_{+}$ for any $ k\in\ccalI$  defined as
\begin{align}
g_{k}(\udV) =    \log \left|\bbI + \sum_{(\ude,q)\in\udV} p_\ude \bbH_k^q \bbW_\ude \bbW_\ude^\dag (\bbH_k^q)^\dag\right|, \label{fnfkt}
\end{align}
which from Proposition \ref{prop:submod1} can be deduced to be a submodular set function over  $\udF$.
From this fact, it follows that the functions $g_{k,\ell}(\udV)=g_k(\udV\cap\udF^\ell),\;\forall\;\udV\subseteq\udF$, for $1\leq \ell\leq L$ are all submodular set functions, where
\begin{align}
\udF^\ell := \left\{(\ude, \ell) | ~\ude \in \udE \}\right\}.
\end{align}
so that $\{\udF^\ell\}_{\ell=1}^L$ form a partition of $\udF$.
Consequently, the set function
\begin{align}
\tilde{g}_{k}(\udV) =    \frac{1}{\Theta(1-\theta_k)} \sum_{\ell=1}^Lg_{k,\ell}(\udV),\;\;\forall\;\udV\subseteq\udF, \label{fnfkt2}
\end{align}
 being a linear combination of submodular set functions in which the combining coefficients are all positive constants, is a submodular set function over  $\udF$.
Next, since truncation preserves submodularity, we can conclude that $f_{k}(\udV)=\min\{\tilde{g}_{k}(\udV),1\},\;\forall\;\udV\subseteq\udF$ is a  submodular set function over $\udF$. Finally, we can expand $f(\cdot)$ in (\ref{fnf}) as
\begin{align}
f(\udV) =   \sum_{k\in\ccalI}  \mu_k f_{k}(\udV),\;\forall\;\udV\subseteq\udF,
\end{align}
which again being a linear combination of submodular set functions (with positive and constant combining weights) is thus a submodular set function over  $\udF$. \myQED
\section{Proof of Propositions \ref{prop:reformD} and \ref{prop:reformD2}}\label{app:reformD}
We will first prove Proposition \ref{prop:reformD}. Here,
 the fact that each $R_k^\tau(.),1\leq k\leq K,\forall\;\tau$ is a monotonic set function over $\udE$ suffices to assert that the problem in (P2D)   can be further constrained without loss of optimality by enforcing that each precoder used must lie in the set $\tilde{\ccalC}_d$. Further,
 without loss of generality, we can assume that each codeword in the given set $\{\breve{\bbW}^{(\ell)}\}_{\ell=1}^L$ is maximal since otherwise the set can always be arbitrarily expanded. Suppose now that an optimal solution involves employing  an identical set of maximal precoding matrix codewords, $\{\hat{\bbW}^{(\ell)}\in\tilde{\ccalC}_d\}_{\ell=1}^L$,  for more than $Q=\lceil\frac{\Theta}{\min_k\Delta_k}\rceil$ scheduling intervals. Consider the first $Q$ scheduling intervals over which the set $\{\hat{\bbW}^{(\ell)}\in\tilde{\ccalC}_d\}_{\ell=1}^L$ is used. Note  that upon using that set over $Q$ scheduling intervals, each user $k$ for whom the accumulated rate is less than $\Theta$ must satisfy
 $\sum_{\ell=1}^LR_k^\ell(\hat{\bbW}^{(\ell)})<\min_j\Delta_j$. As a result, all further uses of the set $\{\hat{\bbW}^{(\ell)}\in\tilde{\ccalC}_d\}_{\ell=1}^L$ can be  replaced without loss of optimality by those of the set $\{\breve{\bbW}^{(\ell)}\in\tilde{\ccalC}_d\}_{\ell=1}^L$, since in any scheduling interval the latter set can simultaneously achieve a larger rate than $\{\hat{\bbW}^{(\ell)}\in\tilde{\ccalC}_d\}_{\ell=1}^L$ for each remaining user. 
 Finally, no more than
  $Q$ uses of the set $\{\breve{\bbW}^{(\ell)}\in\tilde{\ccalC}_d\}_{\ell=1}^L$ are needed to ensure an accumulated rate of at-least $\Theta$ for each user.

We now prove Proposition \ref{prop:reformD2}. Towards this end,
 let us  now construct a matrix $\bbR$ having $K$ rows, one for  each  user.
To build the columns of $\bbR$, enumerate all possible sets of maximal codewords   $\{\bbW^{(\ell)}\in\tilde{\ccalC}_d\}_{\ell=1}^L$ and repeat each set $\lceil\frac{\Theta}{\min_k\Delta_k}\rceil$ times. Next, add a column in $\bbR$ for each such set, where the column contains the rates (in a scheduling interval) achieved upon using that set for all $K$ users.
Clearly, then the sum of each row of $\bbR$ is at-least $\Theta$. Further, after this reformulation upon invoking Proposition \ref{prop:reformD2}, we can deduce that the problem (P2D) {\em is in-fact equivalent to finding a permutation of the columns of the matrix $\bbR$ that minimizes the  weighted sum cover time over the rows, where the cover time of each row is the smallest column index for which the partial sum on that row   is at least $\Theta$}. The latter problem is an instance of the {\em ranking with additive valuations} problem considered in \cite{gamzu_thesis10}. It has been shown in \cite{gamzu_thesis10} that solving a  linear program (LP) followed by a randomized rounding procedure can give rise to a column permutation solution which achieves a weighted sum cover time no greater than a constant times the optimal one. However, the number of constraints in the pertinent LP here grows exponentially with the number of columns in $\bbR$ which requires additional processing to avoid exponential complexity, but still renders this method prohibitively complex. Another deterministic algorithm with a weaker guarantee has also been proposed for the ranking problem \cite{gamzu_soda11}. However a direct adaptation of this algorithm to (P2D) will yield Algorithm \ref{alg:p2DCB} albeit where the maximization in (\ref{p2ppAlgo}) must be optimally solved over $\{\bbW^{(\ell)}\in\tilde{\ccalC}_d\}_{\ell=1}^L$. The latter optimization problem is hard to solve (indeed it is NP-hard) which can dramatically increase the complexity due to the potentially large cardinality $|\tilde{\ccalC}_d|^L$. The key modification introduced in Algorithm \ref{alg:p2DCB} is to sub-optimally and efficiently solve (\ref{p2ppAlgo}) over a larger set $\{\bbW^{(\ell)}\in\ccalC_d\}_{\ell=1}^L$ instead, after recognizing it to be a submodular maximization problem subject to knapsack constraints. Then, since (\ref{p2ppAlgo}) is approximately solved with a constant-factor approximation guarantee by Algorithm \ref{alg:p2greedy}, which we note also returns a set of maximal codewords, a careful verification of the proof in \cite{gamzu_soda11} reveals that Algorithm \ref{alg:p2DCB} retains the $O(\ln(1/\epsilon))$ performance guarantee of the direct adaptation. Indeed, the effect of sub-optimally solving (\ref{p2ppAlgo}) is that the constant $\Gamma$ in Proposition \ref{prop:reformD2} is larger (by a factor $\Omega(1/(2L)^{1/\delta})$) compared to the case when (\ref{p2ppAlgo}) is optimally solved.



\smallskip

\newpage

\newpage

\end{document}